\documentclass[10pt,journal,compsoc]{IEEEtran}
\ifCLASSOPTIONcompsoc
  \usepackage[nocompress]{cite}
\else
  \usepackage{cite}
\fi
\ifCLASSINFOpdf
\else
\fi
\usepackage[cmex10]{amsmath}
\newtheorem{definition}{Definition}
\usepackage{extarrows}
\usepackage{amssymb}
\usepackage{multirow}
\usepackage{graphicx}
\usepackage{pdfcomment}
\usepackage{cite}
\usepackage{booktabs}
\usepackage[ruled,linesnumbered]{algorithm2e}

\usepackage{url}
\newtheorem{theorem}{Theorem}

\begin{document}
\title{Applying Quantum Error-correcting Codes for Fault-tolerant Blind Quantum Cloud Computation}

\author{Qiang Zhao,~\IEEEmembership{}
        John~C.S.~Lui,~\IEEEmembership{Fellow,~IEEE,~ACM}   
        
\IEEEcompsocitemizethanks{\IEEEcompsocthanksitem Qiang Zhao and John C.S. Lui are with the Department of Computer Science and Engineering, The Chinese University of Hong Kong, Hong Kong, China, 999077.\protect\\
E-mail:qiangzhao@cuhk.edu.hk; cslui@cse.cuhk.edu.hk
% \IEEEcompsocthanksitem John C.S. Lui are with Anonymous University.
}% <-this % stops an unwanted space
% \thanks{Manuscript received April 19, 2005; revised August 26, 2015.}
}


\markboth{IEEE Transactions on Dependable and Secure Computing}%
{Shell \MakeLowercase{\textit{et al.}}: Applying Quantum Error-correcting Codes for Fault-tolerant Blind Quantum Cloud Computation}



\IEEEtitleabstractindextext{%
\begin{abstract}
Blind Quantum Computation (BQC) is a delegation computing protocol that allows a client to utilize a remote quantum server to implement desired quantum computations while keeping her inputs, outputs, and algorithms private. However, qubit errors during the quantum computation are important issues that need to be addressed. In this paper, we propose a fault-tolerant blind quantum computation protocol with quantum error-correcting codes to avoid the accumulation and propagation of qubit errors during the computational process. We further propose to apply the concatenated codes in the protocol to improve the error correction performance. More importantly, we quantify the resource consumption of photon pulses by our optimal level concatenation codes. Extensive simulation results show that our scheme not only can improve the preparation efficiency but also reduce quantum resource consumption, which shows a significant improvement to realize a fault-tolerant BQC.
\end{abstract}

}

\maketitle

\IEEEdisplaynontitleabstractindextext

% \fi
\IEEEpeerreviewmaketitle

\IEEEraisesectionheading{\section{Introduction}\label{sec:introduction}}

\IEEEPARstart{Q}{uantum} cloud computing is a promising computing paradigm that can bring revolutionary improvements in communications and computations, as well as deal with tasks that are computationally impossible for traditional computers. Quantum cloud computing is a distributed system composed of quantum computers as servers and simple quantum devices as clients that exchange quantum information and computing instructions through quantum and classical channels. It allows clients to utilize remote quantum computing resources to achieve their desired quantum computation. However, classical communication network protocols can not ensure the security of quantum information and computation. One possible solution to address this problem is to design a Blind Quantum Computation (BQC) protocol for quantum cloud computing. BQC enables a client (say Alice) with limited quantum resources to delegate a computing task to a remote quantum server (say Bob) while ensuring privacy during the computing process~\cite{childs2005secure,arrighi2006blind,broadbent2009universal}. In recent years, a number of BQC protocols have been proposed~\cite{dunjko2012blind,Zhao2017Blind,Zhao2018Finite,Zhao2020fault,kashefi2021classical,shan2021multi}. Among these protocols, Universal Blind Quantum Computation (UBQC) is of particular interest as it not only can guarantee that Alice's inputs, outputs, and quantum algorithms are private to Bob but also only requires Alice to prepare single photon states for easy realization.   

However, in realistic situation, qubits and quantum gates can be easily affected by environmental and device noise, which will cause qubits and quantum gates to have rotational errors. Furthermore, with the increase in quantum computing scale, errors caused by noise have a significant impact on the security and reliability of quantum computing. Currently, it has been always a formidable technical challenge to realize a large-scale quantum computation. Fortunately, Noisy Intermediate-Scale Quantum (NISQ) will be available in the near future, which can help quantum computers with 50-100 qubits (and hopefully a higher number of qubits) to perform tasks that surpass classical computers~\cite{preskill2018quantum}. 

UBQC is composed of the following steps: (a) quantum preparation, (b) quantum measurement, (c) one-way quantum communication, and (d) two-way classical communication~\cite{kashefi2022framework}. In the noisy intermediate-scale UBQC, due to the influence of noise, it is inevitable that errors will occur in qubits and quantum gates in each of the above processes~\cite{chien2015fault,shan2021multi}. Hence, UBQC will have a unique challenge in dealing with qubit errors. In addition, quantum gates need to be fault-tolerant to prevent the accumulation and propagation of errors in the subsequent computation. Therefore, UBQC needs to utilize quantum error-correcting codes~\cite{fujii2015quantum} or other quantum codes~\cite{morimae2012blindtopological,brown2020universal} with error-correcting capability as logical qubits to realize a fault-tolerant blind quantum computation. Since quantum error correction imposes a heavy overhead cost in many qubits and gates to ensure the required fault tolerance, there is a fundamental trade-off between fault tolerance and high resource overhead (the number of required pulses). In this context, this work provides an alternative protocol to implement a fault-tolerant blind quantum computation in NISQ and to reduce quantum resource consumption, in terms of the number of photon pulses. Moreover, our quantum error-correcting scheme can also be applied to other types of BQC and perform fault-tolerant quantum computation. The main contributions of this paper are as follows:
\begin{itemize}
    \item[$\bullet$] We present a remote blind quantum error-correcting code preparation protocol on cluster states to correct qubit errors and improve the successful probability of desired qubits. The encoded circuit is transformed to cluster states, i.e. multi-particle entangled graph states. Then, the server prepares quantum error-correcting codes in the framework of measurement-based quantum computation.
    \item[$\bullet$] Based on the above preparation, we propose a fault-tolerant blind quantum computation protocol with quantum error-correcting codes to avoid the accumulation and the propagation of qubit errors to improve the performance in a delegated quantum computation. Since these quantum error-correcting codes are unknown logical qubits to the server, each of them is able to be considered as a logical unit to take place of the original qubit on the brickwork state to perform a fault-tolerant computation. Furthermore, we provide a demonstration of the $\epsilon$-blindness property in the presence of a malicious server, highlighting the robustness of our protocol even in adversarial noise settings.
    \item[$\bullet$] To achieve higher fault-tolerant performance, we propose to utilize the concatenated stabilizer codes, that is to concatenate the same or different stabilizer codes in a multi-level encoding manner, and to prepare high-quality encoded logical qubits to do fault-tolerant computation. Theoretically, we also derive the lower bound of quantum resource consumption in each level of the concatenation codes, i.e. the number of required pulses.
    \item[$\bullet$] To reduce quantum resource overhead in a fault-tolerant quantum computation, we formulate an optimization model of quantum resource consumption, which can describe the ratio of the number of pulses for concatenation codes to the non-coding case under the condition of having the same probability of successful qubit preparation. When resource consumption reaches the minimum, one can achieve the optimal number of levels in the concatenated stabilizer codes.
\end{itemize}

This paper is organized as follows: In Section \uppercase\expandafter{\romannumeral2}, the background and challenges are introduced. In Section \uppercase\expandafter{\romannumeral3}, we present a remote blind quantum error-correcting codes preparation protocol on cluster state to correct qubit errors, and then propose a fault-tolerant blind quantum computation with quantum error-correcting codes to perform the delegated computing. Furthermore, we also present a proof of achieving blindness for our fault-tolerant protocol. In Section \uppercase\expandafter{\romannumeral4}, the multi-level quantum error-correcting codes are concatenated to reduce the error rate of the generated logical qubits. we derive a lower bound of the number of required pulses to evaluate quantum resource consumption for different levels of concatenation codes. In Section \uppercase\expandafter{\romannumeral5}, we present simulation results to validate the theoretical model of the number of required pulses with concatenated codes, and give an optimal level of the concatenation codes. In Section \uppercase\expandafter{\romannumeral6}, we survey the related works. In Section \uppercase\expandafter{\romannumeral7}, we state our conclusions and future work.

 

\subsection{Background}
BQC protocols can generally be divided into two categories, they are (1) measurement-based BQC, and  (2) circuit-based BQC. In 2005, Childs proposed the first BQC protocol~\cite{childs2005secure} which utilized the quantum circuit model and the quantum one-time pad to realize a secure quantum computation. However, it requires clients to have quantum memory and the ability to perform quantum SWAP operations. Since then, other circuit-based BQC protocols~\cite{arrighi2006blind,aharonov2008fault} have been proposed, but all of these protocols require the client to possess certain quantum capabilities, such as quantum memory, quantum measurements, or quantum computing abilities. In 2009, Broadbent, Fitzsimons, and Kashefi proposed the Universal Blind Quantum Computation (UBQC) protocol~\cite{broadbent2009universal}, which is a type of Measurement-Based Quantum Computation (MBQC)~\cite{raussendorf2003measurement}. This protocol only requires the client to prepare the single photons, and other operations can be delegated to a quantum server, which can greatly reduce the client's burden, i.e. quantum resource requirements. on this basis, other BQC protocols~\cite{li2018blind,quan2021verificable,li2014triple,sheng2015deterministic,sheng2016blind,Zhao2020fault} have been devised and aimed to be as practical as possible. To verify the blindness and correctness of practical BQC, additional protocols~\cite{li2018blind,quan2021verificable} were also proposed to work under the condition of noisy channels or malicious attackers. To make the client (or Alice) as close to the classical computer as possible, multiple-server BQC protocols~\cite{li2014triple,sheng2015deterministic} were proposed based on the shared entanglement states between servers. To tolerate more errors, fault-tolerant BQC protocols~\cite{sheng2016blind,Zhao2020fault} were presented based on the idea of quantum error correction. With the development of quantum technology, UBQC is gradually moving from theoretical research to a more practical deployed setting.  

\begin{figure}[htbp]
	\centering
	\includegraphics[width=3.3in]{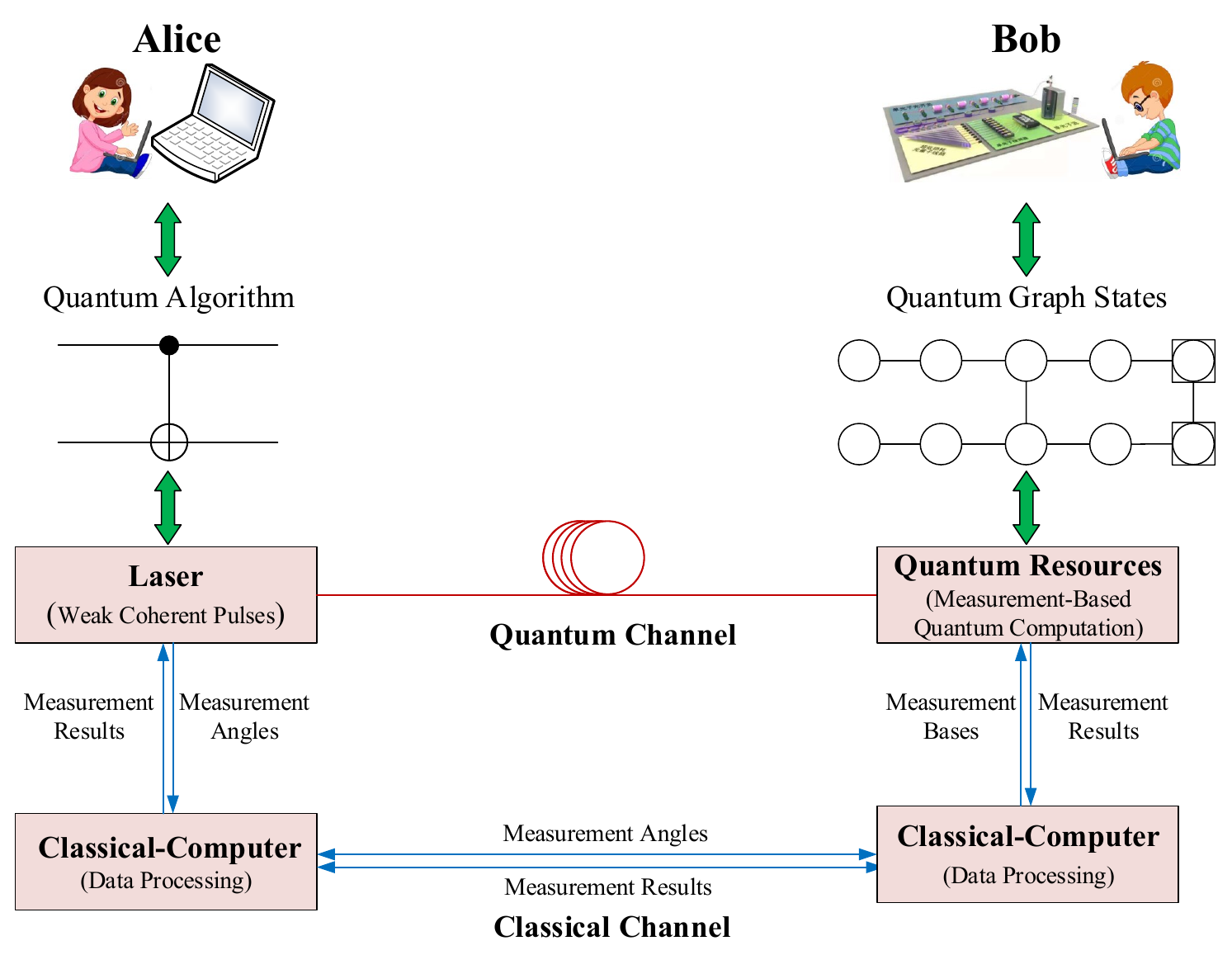}% Here is how to import EPS art
	\caption{The schematic diagram of UBQC.}
\label{fig-schematic-diagram-UBQC}
\end{figure}

In the framework of MBQC, the underlying resource of UBQC is the brickwork states, which are multi-particle entangled graph states constructed from the prepared qubits~\cite{broadbent2009universal,broadbent2010measurement}. A delegated quantum computation can be implemented by measurements on the brickwork state. To illustrate, consider a client (say Alice) and a server (say Bob) with a full-fledged quantum computer. In the preparation stage, Alice prepares a series of single qubits $|+_{\theta}\rangle$ with random polarization $\theta \in \left\{0,\pi/4,...,7\pi/4 \right\}$, and sends them to Bob through a one-way quantum channel, and Bob uses Controlled Z (CZ) gates to act on these received qubits to build a brickwork state. In the interactive measurement stage, Alice meticulously designs the quantum circuit tailored to the computational task at hand, and transforms the sequence of ordered quantum gates in the circuit into corresponding measurement angles, and then sends them to Bob through a two-way classical channel. Bob performs measurement operations on each qubit of the brickwork state, and returns the results to Alice. The measurement procedure is repeatedly performed on each qubit of the brickwork state until the desired quantum computation result is obtained. The schematic diagram of UBQC is shown in Fig.\ref{fig-schematic-diagram-UBQC}.
 
Note that for each qubit in a brickwork state, Alice is able to calculate the measurement angle $\phi _{x,y}^{\prime}$ based on her desired angle $\phi_{x,y}$ and the previous measurement outcomes ~\cite{broadbent2009universal}. In order to ensure the privacy of the computation, the measurement angle $\phi_{x,y}$ needs to be further processed as ${\delta _{x,y}} = \phi _{x,y}^{\prime} + {\theta _{x,y}} + \pi {r_{x,y}},\; {r_{x,y}}{ \in _{R}}\left\{ {0,1} \right\}$. Then, the real measurement angle $\delta_{x,y}$ is sent to Bob to perform the fault-tolerant measurement on each encoded logical qubit ${\left| {{\psi _{x,y}}} \right\rangle _L}$ of the encoded brickwork state. The measurement result $s_{x,y}$ is returned to Alice. If the random bit chosen by Alice is $r_{x,y}=1$, Alice then flips the measurement result $s_{x,y}$; otherwise, she does nothing. 
%Note that each measurement angle sent by Alice is determined by the previous measurement result $\phi$, polarization angle of qubit $\theta$ and random parameter $r$, i.e. $\delta=\phi+\theta+r\pi mod 2\pi$.

\subsection{ Remote blind qubit state preparation protocol with two decoy states}
\label{sec-blind-definition}
For a UBQC, qubits are encoded via the polarization of the single photons using a realistic single-photon source. However, in real implementation, it is possible to send two or more identically polarized photons instead of one, and this may destroy the privacy of the client. Let the ideal joint state shared by the client and the server be described by the state $\pi_{AB}^{ideal}$, and that a malicious server can not infer anything about the client's information. For any UBQC protocol, the ideal joint state $\pi_{AB}^{ideal}$ is evolved as one of the family of states, i.e. $\mathcal{F}(\pi_{AB}^{ideal})$. If the security can be held for any action of the server, any state of the family should be equally blind. In order to analyze the security in a realistic implementation, we consider the settings where the client sends a general state $\rho^{\theta_i}$ instead of a perfect state $|+_{\theta_{i}}\rangle$. The definition of $\epsilon$ blindness is described as follows~\cite{dunjko2012blind}:

\begin{definition}
A UBQC protocol with imperfect preparation is $\epsilon$-blind if the trace distance between the ideal state $\pi_{AB}^{\theta_i}$ and realistic state $\pi_{AB}^{\rho^{\theta_i}}$ is less than $\epsilon>0$. Finally, we have:
\begin{equation}\label{eq-definition-blind}
\mathop {\min }\limits_{\pi _{AB}^{\theta_{i}} \in \mathcal{F}} \frac{1}{2}\left\| {\pi _{AB}^{\{ {\rho ^{\theta _i}}\} } - \pi _{AB}^{\left\{ {{\theta _i}} \right\}}} \right\| \le \varepsilon.
\end{equation}
\end{definition}

To achieve the above security, a Remote Blind qubit State Preparation (RBSP) protocol~\cite{dunjko2012blind} was proposed to prepare $\epsilon$-blind qubits for any security parameter ${\epsilon}>0$. Although one has an ${\epsilon}$-blind UBQC protocol, the client's overhead, which is quantified by the number of pulses needed for preparing a qubit, is usually very large for long-distance communication. Therefore, a modified RBSP with two decoy states was presented by Zhao and Li ~\cite{Zhao2017Blind,Zhao2018Finite} to optimize the number of required pulses. They showed that two decoy states protocol only needs a smaller number of pulses than one decoy state. The RBSP with two decoy states is summarized as follows: 
\begin{enumerate}
    \item Alice generates $N$ weak coherent pulses which contain the signal and two decoy states, and sends them to Bob. 
    \item For each received state, Bob reports to Alice which pulses were received. Based on the reported statistic values, Alice calculates the gains in the signal state and the two decoy states, and determines to continue the protocol if they are within the specified thresholds.
    \item Bob discards the decoy states declared by Alice and separates the remaining signal states into $S$ groups (in fact $S$ is the computational scale). Each group only needs to perform the interlaced 1-D cluster computation subroutine(I1DC)~\cite{dunjko2012blind} so that the desired qubit $|+_{\theta}\rangle$ can be prepared. 
    \item With Alice's knowledge about the angles of signal states in the I1DC procedure and outcomes reported by Bob, Alice can compute the polarization angle $\theta$ of the prepared qubit.
\end{enumerate}

For a UBQC protocol with computation size S, if the client's preparation is $\epsilon$ robust and $\epsilon$ blind for a chosen $\epsilon>0$, the lower bound of the total number of required pulses in RBSP with two decoy states, denoted as $N^{L,v_1,v_2}$, satisfies this following inequality:
	\begin{eqnarray}\label{eq-Low-bound-N}
	N \geq N^{L,v_1,v_2}=\frac{S}{{{p_\mu }{Q_\mu }}}\frac{{\ln \left( {\epsilon /S} \right)}}{{\ln \left( {1 - {p_1^{L,v_1,v_2}}} \right)}},
	\end{eqnarray}
where the average photon number of signal states is denoted as $\mu$, and the average photon number of two decoy states is denoted as $v_1,v_2$. The gain of signal states is $Q_{\mu}$, which is the probability of the detecting event of the signal pulses. The proportion of single photons in the signal states is described as $p_1$, and the lower bound is ${p_1^{L,v_1,v_2}}$. The probabilities of signal and two decoy states chosen by the client are defined as $p_{\mu}, p_{v_1},p_{v_2}$, respectively, with $p_{\mu}+p_{v_1}+p_{v_2}=1$. The detailed discussion is shown in~\cite{Xu2015Blind,Zhao2018Finite}. 

% When $S\gg M_\mu/M_{multi}$, $M_\mu$ is the total number of signal states, $M_{multi}$ is the number of multi-photon signal states, one can estimate Alice's overhead (the total number of required pulses), and the detailed discussion is shown in~\cite{Xu2015Blind,Zhao2018Finite}. We have the following theorem.~\ref{thm-N} for $\epsilon$-blind UBQC.
% \begin{theorem}\label{thm-N}
% 	 For a UBQC protocol with computation size S, if the client's preparation is $\epsilon$ robust and $\epsilon$ blind for a chosen $\epsilon>0$, the lower bound of the total number of required pulses in RBSP with two decoy states, denoted as $N^{L,v_1,v_2}$, satisfies this following inequality.
% 	\begin{eqnarray}\label{eq-Low-bound-N}
% 	N \geq N^{L,v_1,v_2}=\frac{S}{{{p_\mu }{Q_\mu }}}\frac{{\ln \left( {\epsilon /S} \right)}}{{\ln \left( {1 - {p_1^{L,v_1,v_2}}} \right)}}.
% 	\end{eqnarray}
% \end{theorem}

\subsection{Quantum error-correcting codes}
To correct qubit errors, one needs to first establish an error model for quantum computation. Based on the result of~\cite{preskill1998fault}, we know that the evolution of the qubit can be  expressed as a linear combination of four possibilities: (1) no error occurs, (2) the bit flip $\left|0\right\rangle\leftrightarrow \left|1\right\rangle$, (3) the phase flip $|+\rangle\leftrightarrow|-\rangle$, (4) both a bit flip and a phase flip occur. Therefore, the error super-operator ${\cal E}$ is a diagonal in Pauli basis. The error model has the following form
\begin{eqnarray}\label{eq-error-model}
{\cal E}\left( {\left| \psi  \right\rangle \left\langle \psi  \right|} \right) = \sum\limits_{{E_i} \in E} {p\left( {{E_i}} \right)} {E_i}\left| \psi  \right\rangle \left\langle \psi  \right|E_i^\dag,
\end{eqnarray}
where all error $E_i$ are Pauli operators ${E_i} =  \otimes _{j = 1}^nX_j^aZ_j^b,\;a,b \in \{ 0,1\}$, and $p\left(E_{i}\right)$ is the probability for the error $E_{i}$ to occur. In Eq.\eqref{eq-error-model}, we have the normalization condition $E_i^\dag E_i={\textbf{I}},\forall E_i$, and the trace-preserving constraint $\sum\limits_{{E_i} \in E} {p\left( {{E_i}} \right)}=1$. For correcting qubit errors, we need to determine which of these four possibilities occurred, and then correct the error by applying the appropriate Pauli basis.

In the procedure for determining an error syndrome, one needs to make sure that the quantum state can not be destroyed for subsequent quantum computation, and at the same time, its information has to be private. Therefore, one needs to use a suitable quantum code to handle these qubits, which may occur with errors. A popular quantum code is the $[[n,k,d]]$ stabilizer code~\cite{preskill1998fault,nielsen2002quantum,gottesman1997stabilizer}, which can encode $k$ qubits into $n$ qubits, when $n>k$. The parameter $d$ is the distance of the code. If the distance of a code is $d\in \mathbb{N}$, it can correct up to $\lfloor d/2\rfloor$ simultaneous errors from the error set $E$. In quantum error correction, the diagnosis of error is often called the error syndrome measurement, which is increasingly difficult as the number of logical qubits increases in an encoded block. To illustrate, consider a common stabilizer code with seven qubits, i.e. the 7-qubit Steane code [[7,1,3]], which can encode one data qubit in these seven qubits, and can correct the one-qubit error. The encoded logical qubit basis is denoted as $\{|0\rangle_{L},|1\rangle_{L}\}$, and they are Eq.\eqref{eq-logical-qubits}.
\begin{eqnarray}\label{eq-logical-qubits}
\begin{aligned}
{\left| 0 \right\rangle _L} &= \frac{1}{{2\sqrt 2 }}(\left| {0000000} \right\rangle  + \left| {0001111} \right\rangle  + \left| {0110011} \right\rangle\\
&  + \left| {0111100} \right\rangle + \left| {1010101} \right\rangle  + \left| {1011010} \right\rangle\\ 
&+ \left| {1100110} \right\rangle  + \left| {1101001} \right\rangle )\\
{\left| 1 \right\rangle _L} &= \frac{1}{{2\sqrt 2 }}( \left| {1111111} \right\rangle  + \left| {1110000} \right\rangle  + \left| {1001100} \right\rangle \\
& + \left| {1000011} \right\rangle + \left| {0101010} \right\rangle  + \left| {0100101} \right\rangle\\
&+ \left| {0011001} \right\rangle  + \left| {0010110} \right\rangle)
\end{aligned}
\end{eqnarray}

% Note that the code is also a special case of the CSS code, which is an extension of the classical Hamming code in quantum error correction. The code space is the eigenspace of the generators of the code stabilizer $S$, which is a set of $(n-k)$ independent commuting operators $g$. Each codeword $\left|\psi_m\right\rangle$ obeys the eigenvalue equations $\left|\psi_m\right\rangle=g\left|\psi_m\right\rangle, \forall m=0,...,2^{k}-1$.  The stabilizer code can be also described by the generator matrix $G$, which has $2n$ columns and $n-k$ rows. The generator matrix is denoted as $G=(X_{G}|Z_{G})$. Each row in $G$ encodes a generator $g$ of the stabilizer. The column index of $X_{G}$ and $Z_{G}$ labels the qubits. The positions of the 1's in $X_{G}$ indicate the qubits that are acted on by $X$ in the listed generators, and the 1's in $Z_{G}$ indicated the qubits acted on by $Z$. If a 1 appears in the same position in both $X_{G}$ and $Z_{G}$, then the product $Y=ZX$ acts on that qubit. The generator matrix of the [[7,1,3]] is shown as follows~~\cite{nielsen2002quantum}.

% \begin{eqnarray}\label{eq-generator-matrix}
% \begin{aligned}
% &{G_{[[7,1,3]]}}= {({X_G}|{Z_G})_{[[7,1,3]]}}{\rm{ = }}\\
% &\small\left( {\begin{array}{*{20}{c}}
% 	0&0&0&1&1&1&1\\
% 	0&1&1&0&0&1&1\\
% 	1&0&1&0&1&0&1\\
% 	0&0&0&0&0&0&0\\
% 	0&0&0&0&0&0&0\\
% 	0&0&0&0&0&0&0
% 	\end{array}\left|{\begin{array}{*{20}{c}}
% 		0&0&0&0&0&0&0\\
% 		0&0&0&0&0&0&0\\
% 		0&0&0&0&0&0&0\\
% 		0&0&0&1&1&1&1\\
% 		0&1&1&0&0&1&1\\
% 		1&0&1&0&1&0&1
% 		\end{array}} \right.} \right)
% \end{aligned}
% \end{eqnarray}

\begin{figure}[htbp]
	\centering
	\includegraphics[width=3.3in]{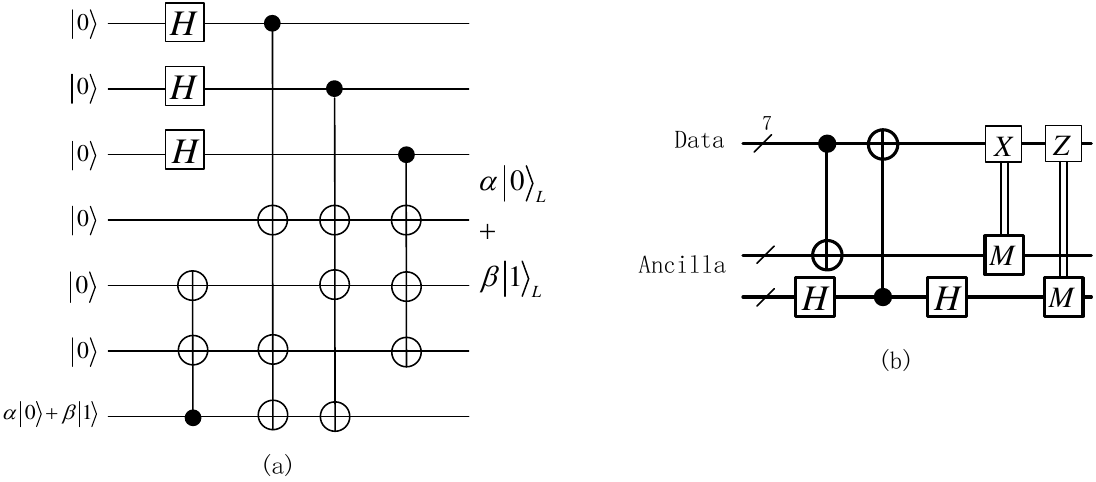}% Here is how to import EPS art
	\caption{The encoding and correction circuits of the [[7,1,3]] code~\cite{preskill1998fault}. (a) An unknown logical qubit and 6 ancilla qubits can be used to encode into [[7,1,3]] code. (b) Two 7-qubit Steane states are used to correct error qubits for a 7-qubit data block. Each CNOT gate in the diagram represents 7 CNOT gates performed in parallel.}
	\label{fig-encode-circuit}
\end{figure}

Note that the [[7,1,3]] encoding circuit of quantum error-correcting codes can be designed as shown in Fig.~\ref{fig-encode-circuit}(a). An unknown logical qubit $\alpha|0\rangle+\beta|1\rangle$ and six ancilla qubits $|0\rangle$ are encoded into $\alpha|0\rangle_L+\beta|1\rangle_L$~\cite{preskill1998fault}. In Fig.~\ref{fig-encode-circuit}(b), two ancilla states are prepared to perform error syndrome measurements of bit flip and phase flip. The measurement results of ancilla qubits are multiplied by the row vectors of $G_{[[7,1,3]]}$ to obtain the parity bits, which can diagnose the error syndrome. Finally, one can use the Pauli operator to correct the qubit errors in the encoded data block.

\subsection{Challenges}
In quantum computation, a basic qubit unit is often stored as the polarization of a single photon or the spin of a single electron. A qubit has the superposition property, which can be described by the two-dimension Hilbert space. A quantum algorithm is described by the quantum circuit which consists of a sequence of quantum gates. In UBQC, the quantum gates are first transformed into the brickwork, and then the server performs measurements on each qubit of the brickwork state to implement the client's computation based on MBQC. Note that quantum gates can be realized by performing measurements on qubits. In a practical UBQC system, one only needs to consider the used qubits. Due to the influence of noise, qubits are prone to polarization errors. Hence, it is necessary to integrate the quantum error correction feature into UBQC to ensure the correctness of a delegated computation. In UBQC, the main processes of quantum computation consist of preparation and measurement. Hence, we state the challenges in terms of the following two processes.

\noindent{\bf \textit{Challenge \uppercase\expandafter{\romannumeral1}:}} In the preparation, we modify the above RBSP protocol to prepare the quantum error-correcting codes instead of single qubits. In order to prepare quantum error-correcting codes, we first design a quantum encoding circuit, and then transform it into graph states. Based on the MBQC model, we can delegate a server to prepare quantum error-correcting codes on graph states. However, both good encoded circuits and good graph states require a large number of data qubits and ancilla qubits, which directly determine the size of the computation. The challenge is to achieve better error correction performance while having a low demand for quantum resources.    

\noindent{\bf \textit{Challenge \uppercase\expandafter{\romannumeral2}:}} In the measurement process, the client sends measurement angles to the server, which uses these angles to perform measurement-based computations on the brickwork state. To prevent the accumulation and propagation of qubit errors in computing, it is crucial to perform fault-tolerant quantum computation. This requires the qubits, quantum gates, and measurements in the computation to be fault-tolerant in the MBQC setting. To achieve this, each qubit of the brickwork state can be encoded as a logical unit using quantum error-correcting code. In this process, one challenge is to ensure the proposed fault-tolerant blind quantum computation protocol is $\epsilon$-blind. Another challenge is to balance the trade-off between the quantum resources consumption and the level of fault tolerance, as a higher fault tolerance often requires a larger number of redundant qubits to counteract the spread of qubit errors.  

\section{Fault-tolerant blind quantum computation with quantum error-correcting codes}

\begin{figure*}[htbp]
	\centering
	\includegraphics[width=1\textwidth]{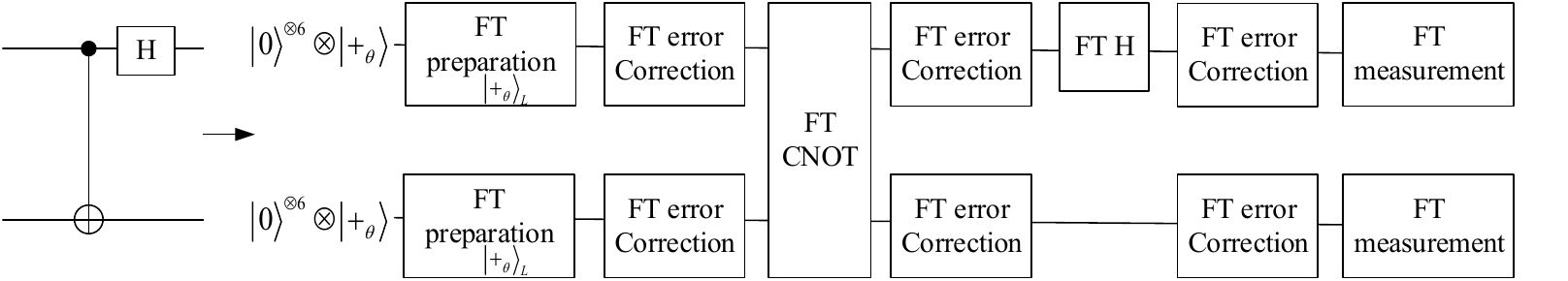}
	\caption{The general process of fault-tolerant quantum computation}
	\label{fig-FTQC-process}
\end{figure*}

To effectively eliminate qubit errors, the encoded quantum gates need to ensure that a failure in executing any encoded gate can only be propagated to a small number of qubits in each encoded data block, so that the aggregated error rate does not exceed the designed fault-tolerant threshold. In addition, since any error correction scheme may introduce errors in the encoded qubits, one must be careful in designing error correction procedures. As a result, we illustrate the following steps to prevent the accumulation and propagation of errors in fault-tolerant quantum computing~\cite{preskill1998fault}: (1) the fault-tolerant preparation of quantum error-correcting codes, (2) fault-tolerant error correction operation, (3) fault-tolerant quantum gates application, and (4) fault-tolerant measurements. This general paradigm is shown in Fig.\ref{fig-FTQC-process}.  

In UBQC, a fault-tolerant computation typically needs to follow the paradigm in Fig.\ref{fig-FTQC-process}. Alice first delegates to Bob to perform the fault-tolerant preparation of quantum error-correcting codes. Then Bob uses these encoded logical qubits to create an encoded brickwork state to realize fault-tolerant quantum computing. In the fault-tolerant preparation, the CNOT gates between non-adjacent qubits are repeatedly used in the fault-tolerant encoded circuit, so a large number of SWAP gates are needed to ensure CNOT gates can act on neighboring qubits, which requires many auxiliary qubits. If these CNOT gates of the encoding circuit can be implemented with high probability, then the error rate of the prepared quantum error-correcting code will be greatly reduced. In the fault-tolerant computation, fault-tolerant quantum gates of the desired quantum circuit need to be equivalently transformed into the brickwork state. Then, the fault-tolerant measurements are performed on each encoded logical qubit of the brickwork state in the framework of MBQC. Consequently, with limited quantum resources, a fault-tolerant blind quantum computation can be implemented by using a combination of quantum error-correcting code preparation and fault-tolerant measurements.

% Hence, it is challenging to achieve the desired fault-tolerant quantum computation with limited quantum resources.
% many qubits are required to prepare quantum error-correcting codes. On one hand, for the delegated preparation, quantum gates of the encoding circuit are equivalently transformed into these brickwork states or other graph states, and then realized by MBQC. On the other hand,
% Since the quantum gates between non-neighboring qubits can not be directly implemented on a brickwork state, one
% needs many SWAP gates to ensure that any quantum gate can act on adjacent qubits.

\subsection{Delegated preparation of quantum error-correcting codes }
In quantum computation, any arbitrary quantum gate can be realized via a cluster state. The essence of a quantum circuit is a series of ordered quantum gates. Hence, each quantum circuit can be implemented via a cluster state~\cite{raussendorf2003measurement}. The encoded circuit can be also realized via a cluster state to generate the encoded logical qubits. These codes can be considered as ''{\em logical units}'' to take the place of the original qubits on a brickwork state to perform fault-tolerant computing. Hence, we propose a remote quantum error-correcting code preparation protocol on a cluster state, which is shown in Protocol.\ref{protocol1-cluster}
\begin{algorithm}[!htb]%[H]%[htbp]
	\caption{A remote quantum error-correcting code preparation on cluster state}
	\label{protocol1-cluster}
	\KwIn {data weak coherent pulses (signal states and two decoy states) with polarization $\rho^{\sigma}$, $\sigma \in_{R} \{ k\pi /4:\;0 \le k \le 7\}$; ancilla pulses with polarization $|+\rangle$.}
	\KwOut {quantum error-correcting codes $\left\{ {|{ + _{{\theta _i}}}\rangle_{L} } \right\}_{1}^S$.}
	
	Alice sends data and ancilla pulses to Bob.
	
	\For{$i$=1 to $S$;}
	{
		Bob prepares qubit $|+_{\theta_i}\rangle$ based on RBSP with two decoy states, and uses it and a group of ancilla qubits $\{|+\rangle\}$ to build cluster state.
		
		\For{$x$=1 to $m$, $y$=1 to $n$;}
		{
			Realization of the quantum encoded circuit on cluster state.
			
			\lIf{$q_{xy}$ is white}{$q_{xy}$ is measured with $\{|0\rangle,|1\rangle\}$ }
			
			\lIf{$q_{xy}$ is green}{$q_{xy}$ is measured with $M(0)$}
			
			\lIf{$q_{xy}$ is red}{$q_{xy}$ is measured with $M(\pi/2)$}			
		}
		Bob prepares the encoded logical qubit $\left\{ {|{ + _{{\theta _i}}}\rangle_{L} } \right\}$.
	}
	\Return  $\left\{ {|{ + _{{\theta _i}}}\rangle_{L} } \right\}$.
\end{algorithm}

% If quantum error-correcting codes are prepared by the fault-tolerant protocol, a large number of ancilla qubits will be needed to design fault-tolerant quantum gates and fault-tolerant measurements. Therefore, it is a formidable task to achieve fault-tolerant quantum computation with limited quantum resources. 

Note that the depth of [[7,1,3]] circuit in Fig.\ref{fig-encode-circuit}(a) is about four, and the most commonly used gate is the CNOT gate. If these CNOT gates can be operated with a high success rate, then the success rate of qubit preparation will also be high. According to the quantum computation model on cluster states~\cite{raussendorf2003measurement}, the quantum gates of [[7,1,3]] circuit can be transformed into a cluster state, and sequentially measure each qubit on the cluster state to prepare the quantum error-correcting code. In the following, we will use the [[7,1,3]] circuit as an example to illustrate how to prepare quantum error-correcting code. 

\begin{figure}[htbp]
	\centering
	\includegraphics[width=3.3in]{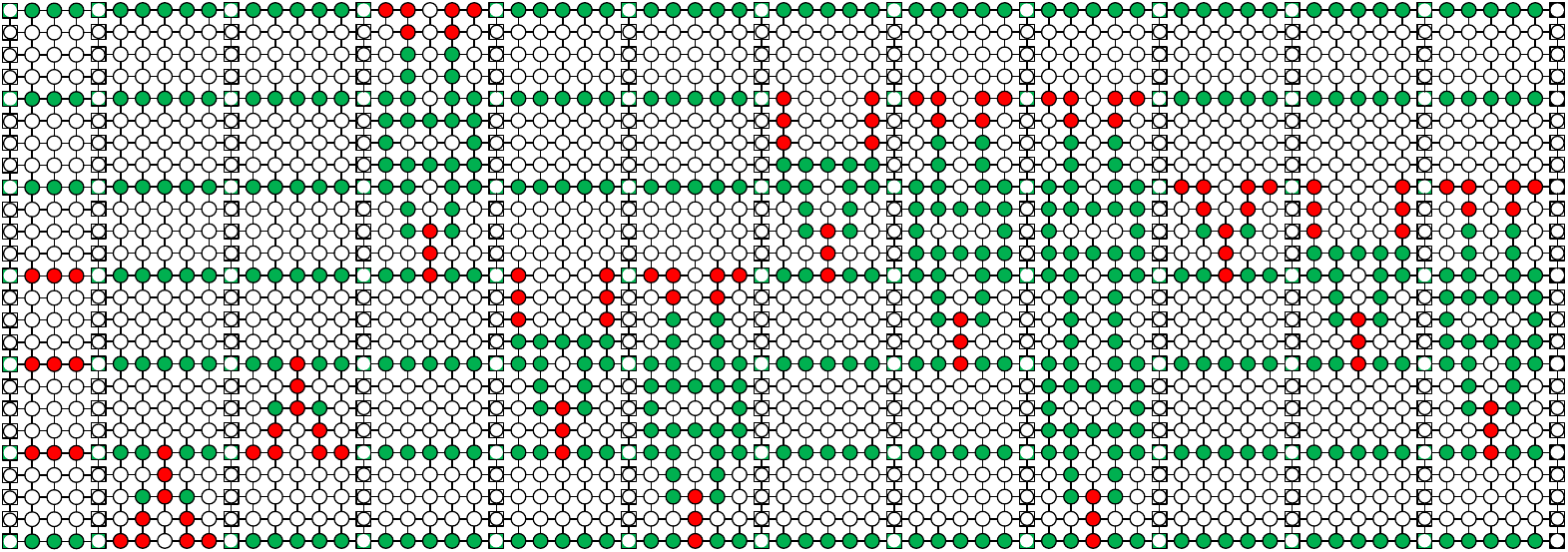}
	\caption{Transformation of the [[7,1,3]] encoded circuit in Fig.\ref{fig-encode-circuit}(a) on cluster state. Circles in green, red and white represent qubits measured in the eigenbasis of $X$, $Y$, and $Z$, respectively.}
	\label{fig-encode-cluster}
\end{figure}

Based on Protocol.\ref{protocol1-cluster}, Alice sends data photon pulses and ancilla photon pulses to Bob. Bob prepares data qubits based on RBSP with two decoy states~\cite{Zhao2018Finite}, and then utilizes a series of ancilla qubits and a prepared data qubit to build the initial cluster state. According to the [[7,1,3]] circuit in Fig.\ref{fig-encode-circuit}(a), Alice can transform each quantum gate into measurement angles on cluster state, as shown in Fig.\ref{fig-encode-cluster}. After that, Bob uses the computational basis $\{|0\rangle,|1\rangle\}$ to eliminate the redundant qubits according to Alice's requirements. The remaining qubits on this cluster state are used to prepare quantum error-correcting code based on Alice's measurement basis $M(\delta)$, which is defined by orthogonal projections on $|\pm_{\delta}\rangle=(|0\rangle \pm e^{i\delta}|1\rangle)/\sqrt{2}$. The parameter $\delta \in [0,2\pi]$ is called the measurement angle. For $\delta=0$ or $\pi/2$, one obtains the $X$ or $Y$ Pauli measurement. The measurement outcome at qubit $i$ will be denoted as $s_{i} \in \mathbb{Z}_2$. We take the specific convention that $s_{i}=0$ if the state collapses to $|+_{\delta}\rangle$, and $s_{i}=1$ if the state collapses to $|-_{\delta}\rangle$.

Note that in the process of eliminating redundant qubits, the structure of the underlying cluster state can be inferred by Bob, and this may cause the leakage of what quantum gates were used. To avoid information leakage, Alice only needs to ensure that, the encoding logical qubit $|+_{\theta_i}\rangle_{L}$ is unknown to Bob during the qubit preparation, while the quantum gates in the encoding circuit can be public. To achieve this goal, one needs to show that the measurement is independent of the polarization angle $\theta$ in the qubit preparation. There are three kinds of measurement bases in the process, including (a) the computational basis $\{|0\rangle,|1\rangle\}$, (b) measurement bases $M(0)$, and (c) $M(\pi/2)$, which correspond to the eigenstates of $Z, X, Y$ in Pauli gates, respectively. These measurement bases are independent of the polarization angle $\theta_i$, which implies that the information of $\theta_i$ will not be leaked to Bob in the preparation process. 

Based on the RBSP with two decoy states, the polarization angle of the prepared qubit $\theta$ is $\epsilon$-blind as defined in Section \ref{sec-blind-definition} which means Bob can not derive any information from it. The prepared quantum state as data qubit will be combined with a series of ancilla qubits to prepare the required quantum error-correcting code $|+_{\theta}\rangle_{L}$ according to Protocol.\ref{protocol1-cluster}. Hence, the quantum error-correcting code is also $\epsilon$-blind.

% the cluster state can be used to prepare the quantum error-correcting code, which is unknown encoding logical qubits to Bob.

\subsection{Fault-tolerant computation of quantum error-correcting codes}

 From the above quantum error-correcting code preparation, we can get that Bob is unable to obtain any information about $\theta$. Hence, these prepared quantum error-correcting codes are all unknown encoded logical qubits to Bob. Each encoded logical qubit can be considered as a ''{\em logical unit}'' to take place of the original qubit on the brickwork state. This new brickwork state can be used to perform fault-tolerant quantum computation. Based on the original UBQC, we propose a novel fault-tolerant blind quantum computation with quantum error-correcting codes, which is shown in protocol.\ref{protocol2-ftbqc}. 

\begin{algorithm}[!htb]%[htbp]
	\caption{Fault-tolerant blind quantum computation with quantum error-correcting codes}
	\label{protocol2-ftbqc}
	\KwIn{data weak coherent pulses with polarization $\rho^{\sigma}, \; \sigma \in_{R} \{ k\pi/4: \; 0 \le k \le 7\}$, and ancilla pulses with polarization $|+\rangle$.}
	\KwOut{measurement result $(s_{x,y})_{1,1}^{nm}$.}
	Alice sends data and ancilla pulses to Bob.
	
	\For{$i$=1 to $S$;}
	{ 
		Bob uses data pulse $\rho^{\sigma}$ to prepare the required qubit ${|{ + _{{\theta _i}}}\rangle}$, and uses it and a group of ancilla qubits $\{|+\rangle\}$ to prepare the desired encoded logical qubit $|{ + _{{\theta _i}}}\rangle_{L}$ based on Procotol.\ref{protocol1-cluster}.		  
	}
	Bob uses these encoded logical qubits $\left\{ {|{ + _{{\theta _i}}}\rangle_{L} } \right\}_{1}^S$ to build the encoded brickwork state.
	
	\For{$x$=1 to $n$; $y$=1 to $m$;}
	{
		Alice calculates angle $\phi _{x,y}^{\prime}$, the beginning values $s_{0,y}^X = s_{0,y}^Z = 0$; and then calculates measurement angle ${\delta _{x,y}} = \phi _{x,y}^{\prime} + {\theta _{x,y}} + \pi {r_{x,y}}$,${r_{x,y}} \in_{R} \left\{ {0,1} \right\}$.
		
		Bob measures encoded logical qubit ${\left| {{\psi _{x,y}}} \right\rangle _L}$ in the base $\{|+_{\delta_{x,y}}\rangle_{L},|-_{\delta_{x,y}}\rangle_{L}\}$, return measurement result $s_{x,y} \in \left\{ {0,1} \right\}$.
		
		\lIf{$r_{x,y}=1$}{flip $s_{x,y}$} 
		
		\lIf{$r_{x,y}=0$}{continue} 
	}
	\Return  $s_{x,y}$.
\end{algorithm}

\begin{figure}[htbp]
	\centering
	\includegraphics[width=3.3in]{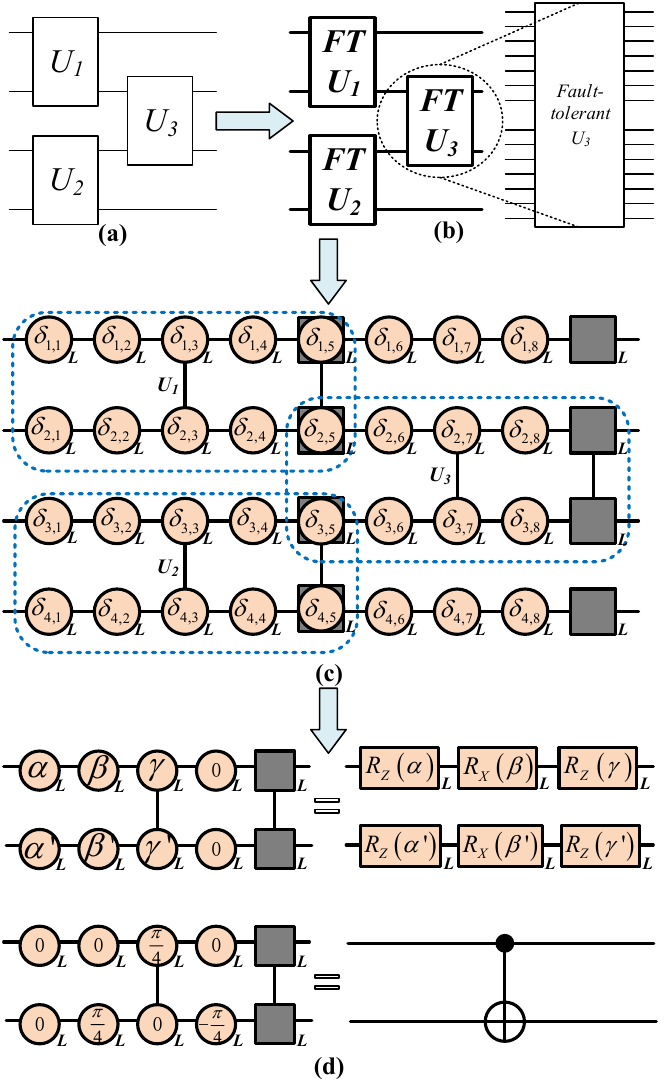}
	\caption{ An example illustration of fault-tolerant blind quantum computation with quantum error-correcting codes. (a)  A simple quantum circuit including three quantum gates $U_1, U_2, U_3$, each thin line represents a qubit. (b) A fault-tolerant quantum circuit, every thick box represents a fault-tolerant quantum gate, and each thick line represents an encoded logical qubit with 7 qubits, i.e. [[7,1,3]] code. (c) Implementation of the fault-tolerant quantum computation on brickwork state, and each angle $\delta$ in circles presents a measurement basis $M(\delta)$, each orange thick circle with subscript L represents an encoded logical qubit, each gray thick box with subscript L represents an output logical qubit. (d) Two brick models built with the encoded logical qubits, i.e. arbitrary rotation logical gate and CNOT logical gate, which can be used to form a universal logic gate group.}
	\label{fig-FTQC-example}
\end{figure}

To illustrate this protocol, consider a simple example as shown in Fig.\ref{fig-FTQC-example}(a). We use our Protocol.\ref{protocol2-ftbqc} to demonstrate the delegated fault-tolerant quantum computation. Firstly, Alice sends weak coherent pulses to Bob, which include data states and ancilla states. The polarization angles of data pulses are selected randomly from $\{k\pi/4:0 \le k \le 7 \}$, and the ancilla is $|+\rangle$. Secondly, Bob uses the data pulse $\rho^{\sigma}$ to prepare the required qubit ${|{ + _{{\theta _i}}}\rangle}$ based on RBSP with two decoy states~\cite{Zhao2018Finite}, and uses it and a group of ancilla qubits $\{|+\rangle\}$ to build cluster state to prepare for the desired encoded logical qubit $|{ + _{{\theta _i}}}\rangle_{L}$, as shown in Fig.\ref{fig-encode-cluster}. Bob uses these encoded logical qubits $\left\{ {|{ + _{{\theta _i}}}\rangle_{L} } \right\}_{1}^S$ to build the encoded brickwork state, as shown in Fig.\ref{fig-FTQC-example}(c). Finally, interactive measurements are performed between Alice and Bob. Alice transforms the quantum circuit to a fault-tolerant quantum circuit, as shown in Fig.\ref{fig-FTQC-example}, and then calculates the angle $\phi _{x,y}^{\prime}$ based on the desired angle $\phi_{x,y}$ and previous measurement outcomes. Let $s_{x,y}^X=\oplus_{i \in D_{x,y}}s_{i}$ be the parity of all measurement outcomes for qubits in $X_{x,y}$, where ${D_{x,y}} \subseteq [x - 1] \times [m]$ is a set with X-dependencies. Similarly, $s_{x,y}^Z = { \oplus _{i \in D_{x,y}^{\prime}}}{s_i}$ is the parity of all the measurement outcomes for qubits in $Z_{x,y}$, where ${D_{x,y}^{\prime}} \subseteq [x - 1] \times [m]$ is a set with Z-dependencies. We assume that the dependency sets $X_{x,y}$ and $Z_{x,y}$ are obtained via UBQC~\cite{broadbent2009universal}. Therefore, the actual measurement angle is $\phi _{x,y}^{\prime}= {\left( { - 1} \right)^{s_{x,y}^X}}{\phi _{x,y}} + s_{x,y}^Z\pi$. The measurement angles $\{\delta_{x,y}\}$ can be calculated according to the original UBQC~\cite{broadbent2009universal}, i.e. ${\delta _{x,y}} = \phi _{x,y}^{\prime} + {\theta _{x,y}} + \pi {r_{x,y}}$, where ${r_{x,y}}$ is randomly chosen in $\left\{ {0,1} \right\}$, as shown in Fig.\ref{fig-FTQC-example}(c). Finally, the fault-tolerant measurements on the encoded brickwork state are used to realize fault-tolerant quantum computing. This procedure is repeatedly performed on the encoded brickwork state until the desired quantum computation result is obtained. Not that arbitrary quantum rotation gate and CNOT gate can be used to build a set of universal logical gates~\cite{deutsch1995universality}. Hence, the fault-tolerant logical gates in the quantum circuit can also be constructed using a series of gates from the set of universal logical gates, which can be used to implement arbitrary fault-tolerant quantum gate, as shown in Fig.\ref{fig-FTQC-example}(d). 

In Protocol.\ref{protocol2-ftbqc}, we note that Alice only uses cluster states to prepare the quantum error-correcting codes according to Protocol.\ref{protocol1-cluster}. In contrast to the original brickwork state, the cluster state can reduce quantum resource consumption. One important note we need to emphasize is that although the encoded circuit in preparation may be leaked to Bob, the prepared quantum error-correcting codes are unknown to Bob. Hence, these codes can be used as ''{\em logical units}'' to build an encoded brickwork state to perform fault-tolerant quantum computation.  

\subsection{Security Analysis}
Based on Protocol.\ref{protocol2-ftbqc}, one can delegate Bob to carry out a fault-tolerant quantum computation. One interesting question is: {\em {If Bob is malicious or dishonest, does Protocol.\ref{protocol2-ftbqc} still ensure the privacy of Alice's information?}} To answer this question, we present the security proof of Protocol.\ref{protocol2-ftbqc} which can ensure that Bob cannot get any computational information except the {\em scale} of the encoded brickwork state (i.e., say number of qubits). Note that in this fault-tolerant computing, Bob only has some information about the measurement bases and the encoded states. Therefore, one needs to show that Alice's information is independent of Bob's information at hand.

\begin{theorem}\label{thm-ftblind}
	The fault-tolerant protocol.\ref{protocol2-ftbqc} is ${\epsilon}$-blind while leaking at most $(n,m)$, where $n$ and $m$ are numbers of rows and columns of the brickwork state respectively, provided that each prepared encoded logical qubit $|+_\theta\rangle_L$ is ${\epsilon}$-blind.
\end{theorem}

\begin{IEEEproof}
	Let $(n,m)$ be the dimension of the brickwork state. According to the definition of brickwork state~\cite{broadbent2009universal}, it ensures that Bob does not get any information on the underlying computation except $n$ and $m$.
	
	Note that on the one hand, one needs to prove that Bob's measurement bases are ${\epsilon}$-independent of Alice's computing information. Alice has two secret bit strings in fault-tolerant quantum computing. One is the polarization angle $\theta$ of quantum error-correcting code, while the other is Alice's computing information, i.e. the actual measurement base $\phi^{\prime}$, which is determined by computational tasks and from previous measurement results. The measurement angle sent to Bob can be calculated according to $\delta=\phi^{\prime}+\theta+\pi r, r \in_{R} \{0,1\}$, and then it is used to perform fault-tolerant measurements on the encoded brickwork state, where $r$ is a random bit string determined by Alice, and it is unknown to Bob. Since each prepared encoded logical qubit is $\epsilon$-blind to Bob, the polarization angle $\theta$ is ${\epsilon}$-uniform in $\{k\pi/4|k=0,1,2,...,7\}$. Hence, $\theta+\pi r$ is ${\epsilon}$-uniform random, and dependent of $\phi^{\prime}$, which implies that Bob's measurement angle $\delta$ is ${\epsilon}$-independent of Alice's computing information $\phi^{\prime}$.
	
	On the other hand, one needs to demonstrate the quantum system for Bob is also ${\epsilon}$-dependent of $\phi^{\prime}$. Since the bit $r_{x,y}$ is random, each encoded logical qubit for Bob will have one of the following two possibilities:
	
	\noindent1)If $r_{x,y}=0$ and $\delta_{x,y}=\phi_{x,y}^{\prime}+\theta_{x,y}$, then $|{\psi _{x,y}}{\rangle _L} = \frac{1}{{\sqrt 2 }}\left( {{{\left| 0 \right\rangle }_L} + {e^{i\left( {{\delta _{x,y}} - \phi _{x,y}^{\prime}} \right)}}{{\left| 1 \right\rangle }_L}} \right)$.
	
	\noindent2)If $r_{x,y}=1$ and $\delta_{x,y}=\phi_{x,y}^{\prime}+\theta_{x,y}+\pi$, then $|{\psi _{x,y}}{\rangle _L} = \frac{1}{{\sqrt 2 }}\left( {{{\left| 0 \right\rangle }_L} - {e^{i\left( {{\delta _{x,y}} - \phi _{x,y}^{\prime}} \right)}}{{\left| 1 \right\rangle }_L}} \right)$.
	
	Without loss of generality, if the angle $\delta$ is fixed, $\epsilon$-blind $\theta$ depends on $\phi^{\prime}$, but the random bit $r_{x,y}$ is unknown to Bob, so the quantum system for Bob is a $\epsilon$-blind mixed state of two cases, which implies it is ${\epsilon}$-independent of Alice's computing information $\phi^{\prime}$. 
\end{IEEEproof}

In summary, using the Protocol.\ref{protocol2-ftbqc}, Bob first can be delegated to prepare the ${\epsilon}$-blind quantum error-correcting codes, and then utilize these codes to perform $\epsilon$-blind fault-tolerant quantum computation. In addition, Protocol.\ref{protocol2-ftbqc} only requires Alice to send weak coherent pulses, which can decouple Alice's dependency on quantum computing and quantum memory. The number of ancilla qubits increases only by a constant factor with the increasing computational scale, rather than a linear increase in the original UBQC~\cite{broadbent2009universal}. Hence, Protocol.\ref{protocol2-ftbqc} has the advantage of reducing Alice's quantum resource consumption and removing Alice's quantum computing requirements. 

\section{Concatenation codes and resource requirements}
\begin{figure}[htbp]
	\centering
	\includegraphics[width=3.3in]{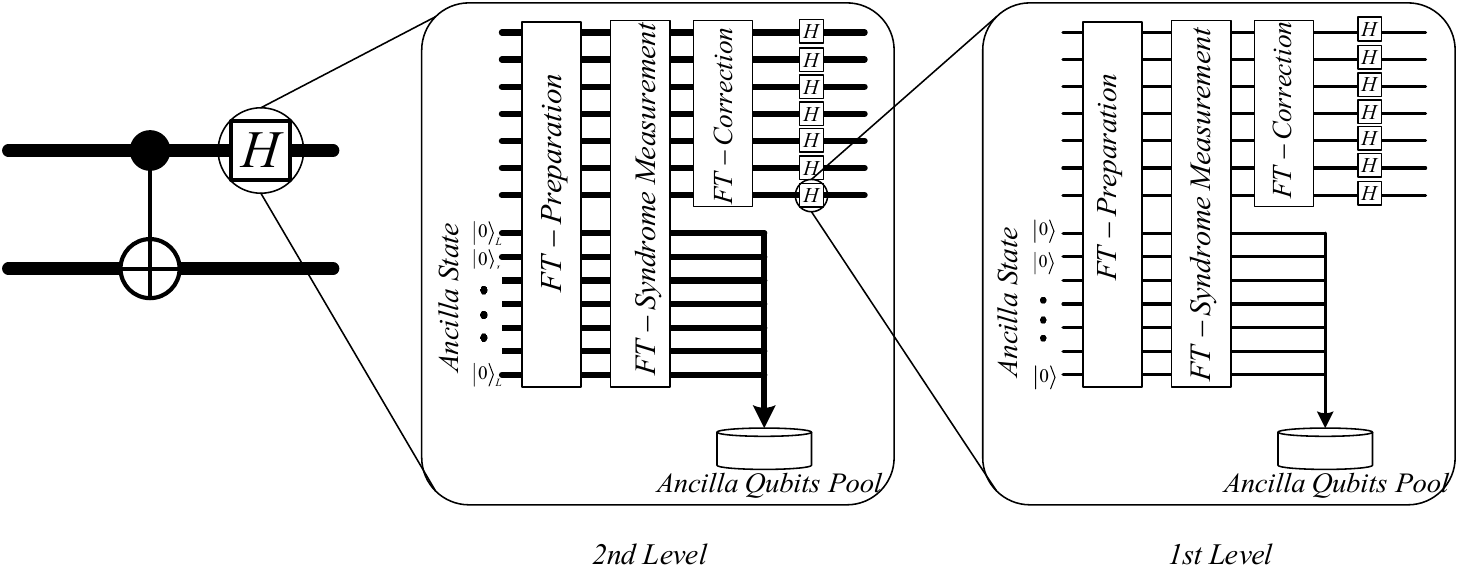}
	\caption{ The two-level concatenated [[7,1,3]] codes. When inspected at higher resolution, each qubit or quantum gate in the block is itself an encoded sub-block. Each block comprises fault-tolerant syndrome measurement, fault-tolerant correction, and fault-tolerant quantum gate(a transversal application of quantum gate).}
	\label{fig-concatenation-code}
\end{figure}

Here, we propose to utilize a concatenated code that further reduces the qubit error rate in quantum computation. The idea of our concatenated code is to recursively execute the above encoding circuit to re-construct a multi-level quantum circuit. When the number of levels is 1, it represents the initial encoded circuit. In the construction, each qubit in the original circuit is encoded in a quantum code whose own qubits are encoded again, and so on, as illustrated in Fig.\ref{fig-concatenation-code}. Hence, in order to improve the error correction performance of qubits, we propose using concatenation codes in UBQC, as shown in Protocol.\ref{protocol3-concatenation}. 

\begin{algorithm}[!htb]%[htbp]
	\caption{Fault-tolerant blind quantum computation with concatenation codes}
	\label{protocol3-concatenation}
% 	\KwIn{data weak coherent pulses with polarization $\rho^{\sigma}, \; \sigma \in_{R} \{ k\pi/4: \; 0 \le k \le 7\}$, and ancilla pulses with polarization $|+\rangle$.}
% 	\KwOut{measurement result $(s_{x,y})_{1,1}^{nm}$.}
% 	Alice sends data and ancilla pulses to Bob.
 \KwIn{Data pulses and ancilla pulses}
 \KwOut{Measurement results $(s_{x,y})_{1,1}^{nm}$} 
 
    \For{$i$=1 to $S$;}
	{
	    Bob uses data pulse $\rho^{\sigma}$ to prepare the qubit ${|{ + _{{\theta _{i,k}}}}\rangle}$ based on RBSP with two decoy states, the initial value $k=0$.\\	    
	  \While{$k=<l-1$;}
	    {
          Bob uses $|{+_{{\theta _{i,k}}}}\rangle_{L}$ and a group of ancilla qubits $\{|+\rangle\}$ to build cluster state, and then prepare the desired encoded logical qubit $|{ + _{{\theta _{i,{k+1}}}}}\rangle_{L}$ based on the $l$th-level concatenated circuit;\\ 
          $k++$.
        }
	    { Output($|{+_{{\theta _{i,l}}}}\rangle_{L}$)}
	    
	}
	Bob uses these encoded logical qubits $\left\{ {|{ + _{{\theta _{i,l}}}}\rangle_{L} } \right\}_{1}^S$ to build the encoded brickwork state.\\	
	The interactive measurement process refers to Protocol.\ref{protocol2-ftbqc} to perform.\\	
	\Return  $s_{x,y}$.
\end{algorithm}

To illustrate our idea, let us consider two-level concatenated [[7,1,3]] codes to reduce the error rate, and each [[7,1,3]] code encodes a single qubit using a block of 7 qubits. When examining one of the 7 qubits in this block at a higher level, we note that it is an encoded sub-block. With this method, the growth of complexity in quantum error correction is not as sharp as the increase in the error-correcting capacity of quantum code. Note that the [[7,1,3]] code can correct one error, so a correction will fail if more than two errors occur in the block. If the probability of error per actual physical qubit at the lowest level is $e_{0}$, and the correction is fault-tolerant, then the probability of the correction failure in [[7,1,3]] code is of order $e_{0}^2$. If we concatenate the code to construct a new block, then an error will occur in the block if two of the sub-blocks of size 7 fail, which occurs with a probability of order $e_{0}^4$. 

%\begin{figure}[htbp]
%\centering
%\includegraphics[width=2.5in]{figures/concatenation-code.eps}% Here is how to import EPS art
%\caption{\label{fig-concatenation-code} This figure shows the two-level concatenated [[7,1,3]] code of a two-gate circuit. The circuit contains a CNOT gate and a Hadamard gate. When examining at high resolution, we can see the next level circuit, i.e. $2$th level. The figure shows only part of it, i.e. the Hadamard gate in the circuit is transformed to its encoded version in the $2$nd level. It comprises fault-tolerant preparation, fault-tolerant syndrome measurement, fault-tolerant correction, and a transversal application of the Hadamard gate. When inspected at higher resolution, we get to see the actual physical circuit, i.e. 1st level. Here we just show the encoding of the Hadamard gate in 1st level.}
%\end{figure}
Let the error probability of each block at level $i$ be denoted as $e_{i}$. At each level of the concatenated code, the block fails if there are errors in at least two contained sub-blocks. For the $i$-level concatenated code, the failure probability can be estimated according to Eq.\eqref{eq-error-pro}.

\begin{eqnarray}\label{eq-error-pro}
\begin{aligned}
{e_i} &= \sum\limits_{k \ge 2}^7 {\left( \begin{array}{l}
	7\\
	k
	\end{array} \right)} e_{i - 1}^k\\
&\approx \left( \begin{array}{l}
7\\
2
\end{array} \right)e_{i - 1}^2 + o\left( {e_{i - 1}^2} \right)\\
&\approx \frac{{{{\left( {21{e_0}} \right)}^{{2^i}}}}}{{21}} + o\left( {e_0^{{2^i}}} \right).
\end{aligned}
\end{eqnarray}

In order to significantly reduce the probability of error, the condition $e_0 < 1/21$ must be satisfied in the concatenation codes (the concatenation codes hold if and only if $(21e_0)^{2^i}\ll1$). Further, we can also make the error rate arbitrarily small by adding sufficient levels of concatenation. If the concatenated [[7,1,3]] code circuit is delegated to Bob to prepare quantum error-correcting code on the cluster state in Protocol.\ref{protocol2-ftbqc}, then the number of required ancilla pulses will increase with the number of levels. Since a data qubit is used to prepare an encoded logical qubit using [[7,1,3]] encoding circuit, the number of required data qubits is consistent with that of the non-coding case, which remains unchanged.
%\begin{eqnarray}\label{eq-FT-pro}
%\begin{aligned}
%{e_{i + 1}} = \sum\limits_{k \ge 2}^7 {\left( \begin{array}{l}
%7\\
%k
%\end{array} \right)e_i^k}  = {\left( {{e_i} + 1} \right)^7} - \left( {1 + 7{e_i}} \right).
%\end{aligned}
%\end{eqnarray}

\begin{figure}[htbp]
	\centering
	\includegraphics[width=2.5in]{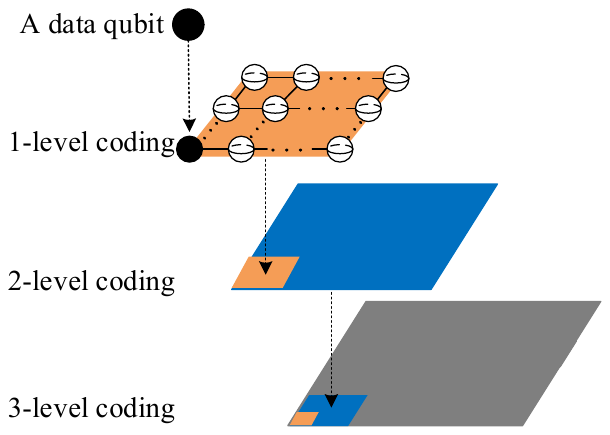}
	\caption{The encoding structure for a date qubit in 3-level concatenation code.}
	\label{fig-multi-level-ancilla}	
\end{figure}

Note that in Protocol.\ref{protocol3-concatenation}, weak coherent pulses are sent to Bob, which contain data pulses and ancilla pulses. According to RBSP with two decoy states, data pulses are used to prepare the required data qubits $\{|+_{\theta_i}\rangle\}_1^S$, and each data qubit is encoded into a required logical qubit. Hence, the number of data pulses is equal to RBSP with two decoy states, which is denoted as $N^{d} $~\cite{Zhao2020fault}. The number of ancilla pulses depends on the encoding circuit on cluster state, as shown in Fig.\ref{fig-encode-cluster}. Note that the required number of ancilla qubits is a constant when preparing an encoded logical qubit, and we denoted it as $C$. If the transmittance of the  quantum channel is $T$, and the computational scale is $S$, then the number of ancilla pulses satisfies $N^{a}=CS/T$. Hence, the total number of required pulses in Protocol.\ref{protocol3-concatenation} is the sum of data and ancilla pulses, i.e.$N=N^{d}+N^{a}$.

The ancilla pulses are used for preparations, syndrome measurements, and error corrections. Note that, the required ancilla qubits in the syndrome measurements and corrections can be reused in our fault-tolerant protocol, so this part of the resource consumption of ancilla qubits can be ignored, and only the number of ancilla qubits in the preparation is needed for analysis. In Protocol.\ref{protocol3-concatenation}, we use cluster state to prepare the encoded logical qubits. Based on the circuit model, the block size of ancilla qubits to prepare an encoded logical qubit is a constant $C$. According to the encoding structure for data qubits in Fig.\ref{fig-multi-level-ancilla}, the block size of ancilla qubits is the sum of the required number at all levels. If we use an $n$-level concatenation code to prepare an encoded logical qubit, the number of ancilla qubits is denoted as $N_n^a$, and its value can be calculated as follows:

\begin{equation}
N_n^a=\sum\limits_{i = 1}^n {{7^{i - 1}}C}  = \frac{\left( {{7^n} - 1} \right)C}{6}.
\end{equation} 
The number of required data pulses is consistent with RBSP with two decoy states. According to Eq.\eqref{eq-Low-bound-N}, we have $N^{d}\ge N^{L,v_{1},v_{2}}$. The required total number of pulses in the $n$-level concatenated [[7,1,3]] code, i.e. $N_{n}$, is estimated as follows:
\begin{eqnarray}\label{eq-FT-Nn}
\begin{aligned}
N_n &={N^{d}}+{N_{n}^{a}}\\
&\ge {N^{L,{v_1},{v_2}}} +\frac{{\left( {{7^n} - 1} \right)CS}}{{6T}}\\
&\approx \frac{S}{T}\left[\frac{{\ln \left( {\epsilon/S} \right)}}{{{p_\mu }\mu \ln \left( {1 - p_1^{L,{v_1},{v_2}}} \right)}} + \frac{{\left( {{7^n} - 1} \right)C}}{{6}}\right],
\end{aligned}
\end{eqnarray}
where $S$ is the computational scale, $\epsilon$ is the security parameter of the blind quantum computation, $T$ is the transmittance of the quantum channel between Alice and Bob, $\mu$,$v_1,v_2$ are the average number of signal photons and two decoy photons, respectively, the proportion of single photon in the signal states is described as $p_1$, and the lower bound is ${p_1^{L,v_1,v_2}}$, the probability of signal pulses chosen by the client is defined as $p_{\mu}$.

In the RBSP protocol with two decoy states, for the computation size $S$, if an error occurs in the prepared qubits, the whole preparation protocol will fail. We assume that the preparation protocol is a repeatable Bernoulli experiment. If the error probability of each prepared qubit is $e_{0}$, the probability of successful qubit preparation in each experiment will be $(1-e_{0})^S$. After repeating $k$ times, the probability of successful qubit preparation will be $1-(1-(1-e_{0})^S)^{k}$. In the $n$-level concatenated code, the probability of successful qubit preparation is $(1-e_{n})^S$. If the probability 
 of successful qubit preparation is the same in both coding and non-coding cases, i.e. $1-(1-(1-e_{0})^S)^{k}=(1-e_{n})^S$, then the repetition number $k$ can be calculated as follows:
\begin{equation}\label{eq-repeat-k}
k = \ln [1 - {(1 - e_{n})^S}]/\ln [1 - {(1 - e_{0})^S}], 
\end{equation}
where $e_{n}$ can be estimated according to Eq.\eqref{eq-error-pro}. In other words, if Alice wants to use the non-coding protocol to achieve the same probability as encoding, the number of required data pulses is $k$ times that of the previous case, i.e. $kN^d$. Let $R$ describe the ratio of the number of pulses for concatenation codes to that of the non-coding case at the same probability. The ratio $R$ is a resource consumption function with the number of levels $n$, which can be used to estimate the optimal level of concatenated codes. The resource consumption ratio $R(n)$ is shown as follows:
\begin{eqnarray}\label{eq-n-level}
\begin{aligned}
R\left( n \right) &= \frac{{{N_n}}}{{k{N^d}}} = \frac{{{N^d}{\rm{ + }}N_n^a}}{{k{N^d}}} \le \frac{{{N^{L,{v_1},{v_2}}} + N_n^a}}{{k{N^{L,{v_1},{v_2}}}}}\\
&= \frac{{\ln \left[ {1 - {{\left( {1 - {e_0}} \right)}^S}} \right]}}{{\ln \left[ {1 - {{\left( {1 - {e_n}} \right)}^S}} \right]}}\left( {\frac{{\left( {{7^n} - 1} \right)CS}}{{6T{N^{L,{v_1},{v_2}}}}}{\rm{ + }}1} \right),
\end{aligned}
\end{eqnarray}
where $N^{L,v_1,v_2}$ is the lower bound of the number of required pulses in the RBSP with two decoy states, which can be estimated according to Eq.\eqref{eq-Low-bound-N}. When the computation scale $S$ is fixed, we can obtain the ratio $R\left( n \right) \sim {\left( {7/2} \right)^n}$ according to Eq.\eqref{eq-error-pro} and Eq.\eqref{eq-n-level}. Hence, the ratio $R$ has an exponential growth trend for a large number of levels $n$. For a high-level concatenation code, a large number of qubits are used for repeated encoding. In our concatenation code, we not only need to consider improving the probability of successful qubit preparation but also reducing resource consumption. Hence, we only take into account the optimal level in the low-level concatenated code ($0\le n\le4$). When the resource consumption ratio $R(n)$ reaches its minimum, it indicates that the utilization efficiency of pulses is at its maximum compared to the non-coding case, which implies that the performance of concatenated code is optimal at this point. 

% At this point, only a small number of pulses for concatenation code can be used to achieve a highly successful probability, in which case the corresponding level $n$ is the optimal one.  

In summary, according to Eq.\eqref{eq-FT-Nn}, note that the number of required data pulses is unchanged with the increasing number of levels, and the number of required ancilla pulses grows exponentially with the increasing number of levels in the concatenation code. Combined with Eq.\eqref{eq-n-level}, we can estimate the resource consumption ratio $R(n)$, and further determine the optimal level $n$. 
%in practice, it is impossible to avoid the error qubits in the transmission and measurement because of the noises and imperfect devices. Hence, we need to design a fault-tolerant quantum circuit.

\section{Performance Evaluation}
In this section, we present the performance and simulation results of our protocols to demonstrate the advantages of quantum error-correcting codes in fault-tolerant blind quantum computation.

\noindent{\bf {Simulation settings:}} Our computer uses Matlab with an Intel(R) Core(TM) i7-6700HQ CPU with 12.0GB RAM and Win 10 pro OS to carry out these simulations. The overall transmittance $T$ (including fiber transmittance and detection efficiency) is calculated as follows:
\begin{equation}
T = {t_s}\cdot{\eta _s}\cdot{10^{ - \alpha L/10}}
\end{equation}

\begin{table}[htbp]
	\centering
	\caption{Simulation parameters for our protocol}
	\label{table-parameter}
	\setlength{\tabcolsep}{1mm}{
		\begin{tabular}{ccccccccccccc}
			\toprule[1.5pt]
			$\alpha$ & $t_{S}$ & $\eta_{S}$  & $\mu$ & $v_{1}$ & $v_{2}$ & $p_{\mu}$ & $p_{v_1}$ & $p_{v_2}$ & $S$ & $\epsilon$ & $e$\\
			\midrule[1pt]
			0.2 & 0.45 & 0.1 & 0.6 & 0.125 & 0 & 0.9 & 0.05 & 0.05 &  1000 & $10^{10}$ & 0.01\\
			\bottomrule[1.5pt]
	\end{tabular}}
\end{table}
Table.\ref{table-parameter} (please refer to the data in~\cite{ma2005practical} and~\cite{Xu2015Blind}) summarizes our simulation parameters, where the parameter $\alpha$ is the loss coefficient measured in dB/km, $L$ is the length of the fiber in km, $t_{s}$ is denoted as the internal transmittance of optical components on Bob's side, and ${\eta_{s}}$ is the detector efficiency on Bob's side. The mean photon number of signal states and two decoy states are represented as $\mu, v_1, v_{2}$, respectively. The chosen probability of signal states and two decoy states are denoted as $p_{\mu}, p_{v_1}, p_{v_2}$, respectively. The computation scale is $S$. The required blindness of the UBQC is denoted as $\epsilon$. The error probability of each qubit is denoted as $e_0$. Based on \ref{fig-encode-cluster}, we calculate the number of ancilla qubits, $C=1774$. 

We assume that Alice has a laser transmitter with frequency $f=1MHZ$, and Bob has a full-fledged quantum computer. According to the setting in Table.\ref{table-parameter}, one can derive the preparation efficiency of our protocols, that is the number of qubits generated per second. Combined with Eq.\eqref{eq-FT-Nn}, the upper bound of the efficiency in the concatenation code is estimated as.
\begin{equation}\label{eq-efficiency}
\begin{aligned}
E &= S \cdot f/{N_n}\\
&\le S \cdot f/\left( {{N^{L,{v_1},{v_2}}} + N_n^a} \right)
\end{aligned} 
\end{equation}

% \noindent {\bf A. simulation \uppercase\expandafter{\romannumeral1}: The number of required pulses for preparing [[7,1,3]] codes} 

\noindent {\bf A. simulation 1: Performance evaluation with [[7,1,3]] codes} 

\noindent {\bf Purpose of simulation 1}: In Protocol.\ref{protocol1-cluster}, we utilize the cluster state to prepare quantum error-correcting codes for correcting error qubits. In order to show the advantages of our approach, we perform simulations to evaluate the quantum resource consumption and preparation efficiency.

\begin{figure}[htbp]
	\centering
	\includegraphics[width=2.5in]{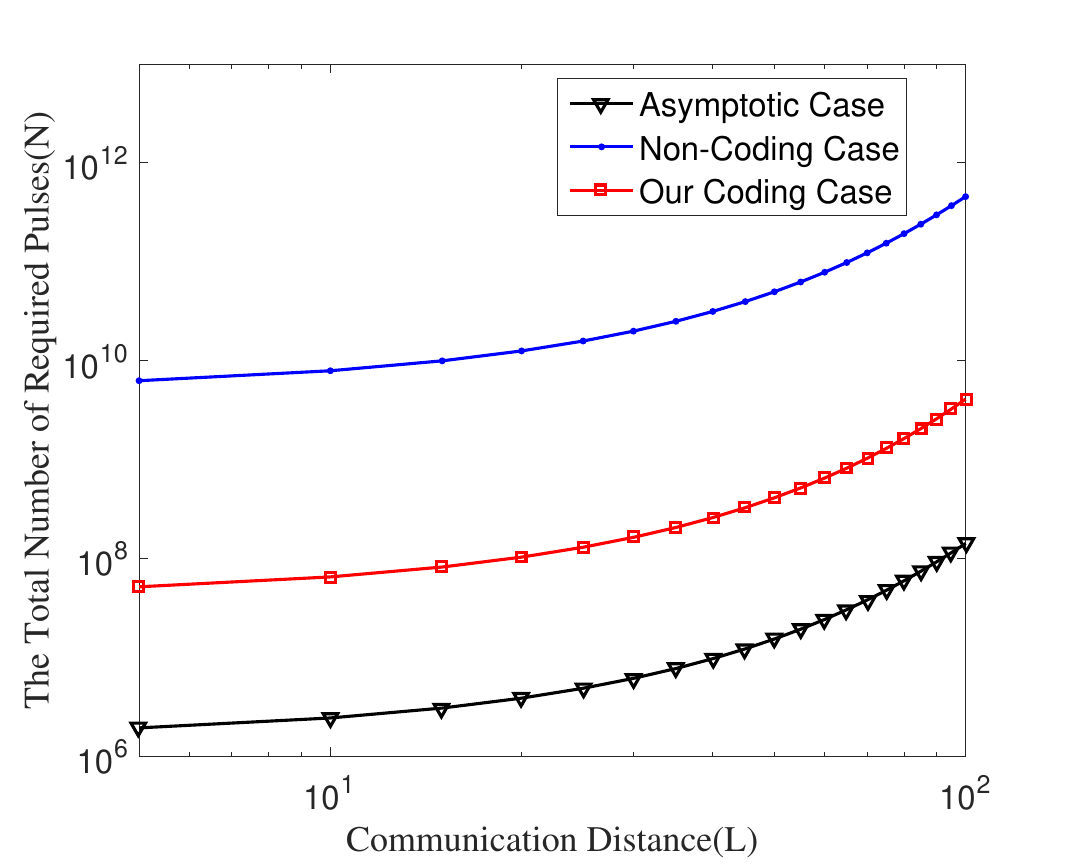}
	\caption{ The total number of required pulses $N$ with the same probability of successful qubit preparation vs. the communication distance between Alice and Bob. 
 % The green and red lines show the simulation results of our Protocol.\ref{protocol1-cluster} with the non-coding case and the coding case, respectively. The black line shows the simulation results in the asymptotic case (infinite data size and with near-perfect qubits preparation).
 }
	\label{fig-one-Coding-LvsN}
\end{figure}

\noindent {\bf Discussions of simulation 1}:
Fig.\ref{fig-one-Coding-LvsN} shows the relationship between the total number of required pulses and the communication distance under the condition of having the same probability of successful qubit preparation. The red line represents the simulation results of our Protocol.\ref{protocol1-cluster} with our coding case, and the green line represents the non-coding case. The black line represents the simulation results in the asymptotic case which is the infinite data size and with near-perfect qubits preparation~\cite{ma2005practical,Zhao2018Finite}.

From Fig.\ref{fig-one-Coding-LvsN}, one can see that the total number of required pulses for the coding case in Protocol.\ref{protocol1-cluster} is less than that of the non-coding case in RBSP with two decoy states under the condition of having the same probability of successful qubits preparation. Since the initial error probability of each qubit is $e_0$, the probability of successful qubit preparation with size $S$ is $(1-e_0)^S$. In quantum error-correcting code preparation, the error probability of each encoded logical qubit is $e_0^2$, and the probability of successful encoded qubit preparation is $(1-e_0^2)^S$. In order to obtain the same probability of successful qubit preparation, the non-coding case needs to be repeated $k$ times. As the communication distance increases, the value of $N$ grows, which implies that the channel loss and qubit error rate have a significant influence on $N$. For long-distance communication, the advantages of our Protocol.\ref{protocol1-cluster} are more apparent than the non-coding case, which means that the number of required pulses $N$ is closer to the asymptotic case.

\begin{figure}[htbp]
	\centering
	\includegraphics[width=2.5in]{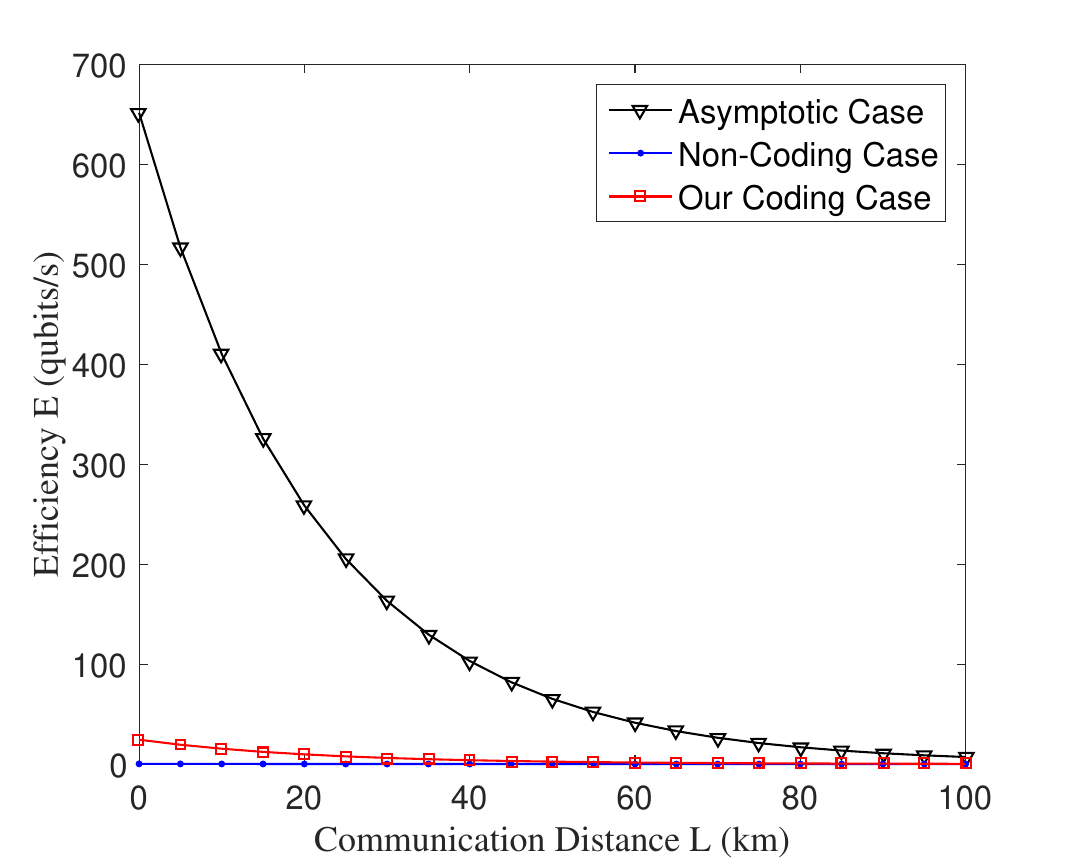}
	\caption{ The preparation efficiency $E$ with the same probability of successful qubit preparation vs. the communication distance $L$. 
 % The green, red, and black lines show the efficiency curves of our protocol.\ref{protocol1-cluster} with coding, non-coding, and asymptotic case respectively.
 }
	\label{fig-one-Coding-LvsEfficiency}
\end{figure}

% \noindent {\bf (b) The efficiency for preparing [[7,1,3]] codes} 
Fig.\ref{fig-one-Coding-LvsEfficiency} illustrates the correlation between the preparation efficiency $E$ and the communication distance $L$ under the condition of having the same probability of successful qubit preparation. The red line in the figure represents the efficiency curves of our Protocol.\ref{protocol1-cluster} in the coding case, while the green and black lines represent the non-coding case and the asymptotic case (infinite data size and with near-perfect qubits preparation), respectively. 

From Fig.\ref{fig-one-Coding-LvsEfficiency}, one can observe that the efficiency of preparing qubits gradually decreases with increasing communication distance. Compared with the non-coding case, the preparation efficiency of qubits for the coding case in our Protocol.\ref{protocol1-cluster} is closer to the asymptotic limit under the same condition. As the error rate decreases from $e_0$ to $e_0^2$, more pulses are used as ancilla qubits to prepare quantum error-correcting codes, which will lead to a drop in efficiency.

\noindent
{\bf Summary of simulation 1:}
We show that the performance with [[7,1,3]] code is better than the non-coding case in terms of both quantum resources and preparation efficiency. Hence, our protocol.\ref{protocol1-cluster} is practical for preparing the required qubits.

\noindent {\bf B. simulation 2: Discussions on concatenation levels}

\noindent {\bf Purpose of simulation 2:} In Protocol.\ref{protocol2-ftbqc}, we employ the quantum error-correcting codes as "{\em logical units}" to facilitate fault-tolerant blind quantum computation. However, the propagation and accumulation of logical qubit errors may result in a relatively high fault-tolerant threshold, Therefore, we propose to use the concatenated [[7,1,3]] code to reduce the error rate of encoded logical qubits, thereby improving the fault-tolerant threshold. The naive implementation of concatenated codes will result in a large number of ancilla qubits. To enhance fault tolerance with limited quantum resources, we perform simulation 2 to identify the optimal level of concatenation.

\begin{figure}[htbp]
	\centering
	\includegraphics[width=2.5in]{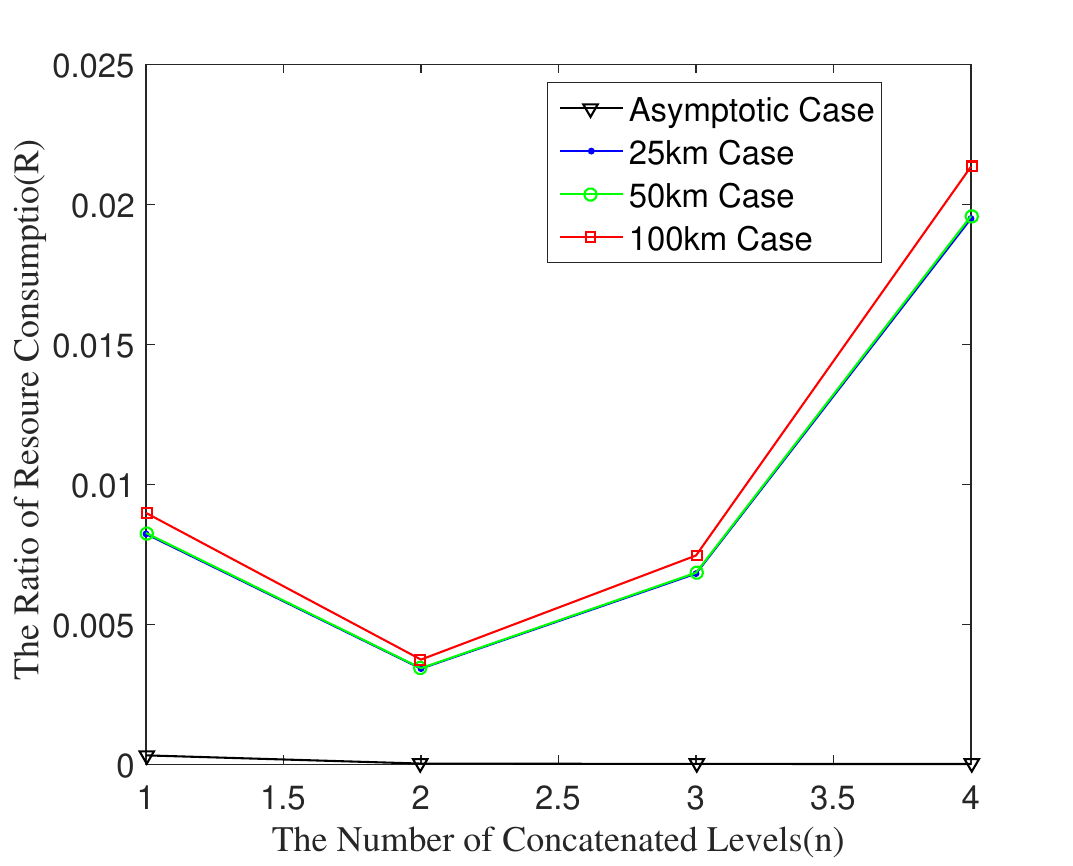}
	\caption{The ratio of resource consumption ($R(n)$) vs. the number of levels ($n$) in the concatenation code. 
 % The blue, green, and red lines show the simulation results of the resource consumption ratio at communication distances of 25km, 50km, and 100km, respectively. The black line shows simulation results in the asymptotic case.
 }
	\label{fig-Ratio-resource-consumption}	
\end{figure}

\noindent {\bf Discussions of simulation 2:}
Fig.\ref{fig-Ratio-resource-consumption} illustrates the relationship between the ratio of resource consumption ($R(n)$) 
and the number of levels ($n$) in the concatenation code. The blue, green, and red lines represent the simulation results of the resource consumption ratio for communication distances of 25km, 50km, and 100km, respectively. The black line corresponds to the simulation results of the asymptotic case, which is the infinite data size and with near-perfect qubits preparation.

According to Eq.\eqref{eq-n-level}, the resource consumption ratio $R(n)$ is the proportion of the number of pulses for concatenated [[7,1,3]] codes in that of the non-coding case. In order to identify the optimal level, one can set the partial derivative $\partial R\left( n \right)/\partial n = 0$ to solve for the extreme value. Nevertheless, the partial derivative is too complicated that it is difficult to solve its analytical solution. Instead, we use simulation to estimate the optimal level $n$, as it is shown in Fig.\ref{fig-Ratio-resource-consumption}. Observe that when the optimal number of levels $n$ is about $2$, the resource consumption ratio $R(n)$ reaches the minimum value. This implies the 2-level concatenated code only needs relatively fewer quantum resources than other levels in meeting the requirements of qubit error rate. As a result, the 2-level concatenated [[7,1,3]] code enables quantum computation with limited resources to achieve better error-correcting performance.

\noindent
{\bf Summary of simulation 2:}
When the concatenated code is used in Protocol.\ref{protocol2-ftbqc}, we discuss the effect of the concatenation levels on the quantum resource consumption under the condition of having the same error rate of qubits. By analyzing the ratio of resource consumption $R(n)$, one can determine that the optimal concatenation level is n=2, which can minimize resource consumption while still maintaining a high level of fault tolerance in UBQC.

\noindent{\bf C. simulation 3: Comparison with concatenated [[7,1,3]] codes}

\noindent{\bf Purpose of simulation 3:}
In Protocol.\ref{protocol3-concatenation}, we propose to use the concatenated codes to perform fault-tolerant blind quantum computation. To further evaluate the performance of concatenated [[7,1,3]] code, we perform simulation 3 to compare the quantum resource consumption and preparation efficiency of the concatenation codes at different levels.

\begin{figure}[htbp]
	\centering
	\includegraphics[width=2.5in]{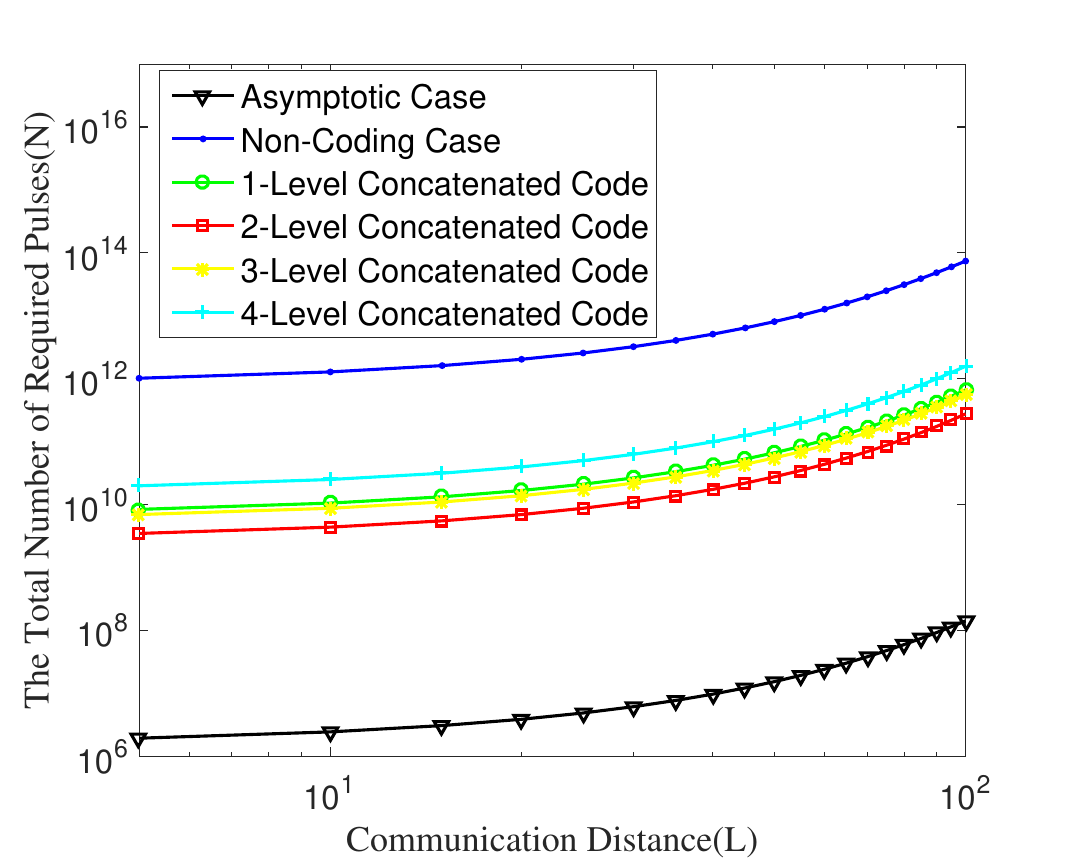}
	\caption{ The total number of required pulses $N$ with the same probability of successful qubit preparation vs. the communication distance ($L$). 
 % The green, red, yellow, cyan, and blue lines show the simulation results of our protocol with the 1,2,3,4-level concatenation code and non-encoding case, respectively. The black line shows the simulation results in the asymptotic case.
 }
	\label{fig-Concatenation-LvsN}
\end{figure}

\noindent{\bf Discussions of simulation 3:} Fig.\ref{fig-Concatenation-LvsN} depicts the relationship between the total number of required pulses $N$ and the communication distance ($L$) under the condition of having the same probability of successful qubit preparation. The green, red, yellow, cyan, and blue lines represent the simulation results of the 1,2,3,4-level concatenation code and non-encoding case, respectively. The black line represents the simulation results in the asymptotic case which is the infinite data size and with near-perfect qubits preparation.

From Fig.\ref{fig-Concatenation-LvsN}, the number of required pulses $N$ in each case is an increasing trend with the communication distance $L$, which indicates that both the channel loss and qubit error rate have a significant influence on $N$. Compared with the non-coding case under the same condition, the number of required pulses $N$ for encoding is closer to the asymptotic case. The reason is that the error probability of the encoded logical qubits prepared by the concatenated circuit is much lower than that of the non-coding case. In order to obtain the same error probability of prepared qubits, the preparation protocol for the non-coding case needs to be repeated $k$ times, which results in wasting a vast number of pulses. Further, one can see that the number of required pulses in the 2-level concatenation code is less than other levels at the same probability of successful qubit preparation, which shows the 2-level concatenation code can provide superior error-correcting performance using limited quantum resources. 

\begin{figure}[htbp]
	\centering
	\includegraphics[width=2.5in]{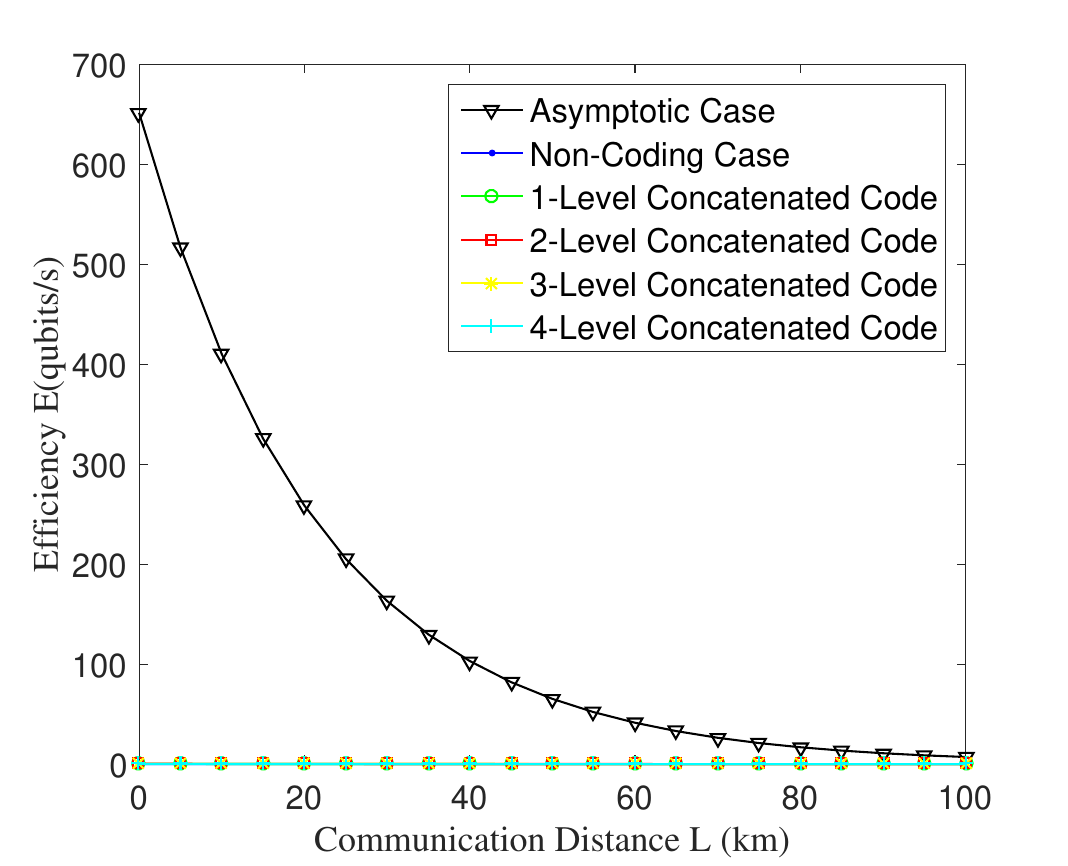}
	\caption{ The preparation efficiency $E$ with the same probability of successful qubit preparation vs. the communication distance $L$. 
 % The green, red, yellow, cyan, and blue lines show the simulation results of our Protocol.\ref{protocol3-concatenation} with 1,2,3,4-level concatenation code and non-coding case, respectively. The black line shows simulation results in the asymptotic case.
 }
	\label{fig-Concatenation-LvsEfficiency}
\end{figure}

\begin{figure}[htbp]
	\centering
	\includegraphics[width=2.5in]{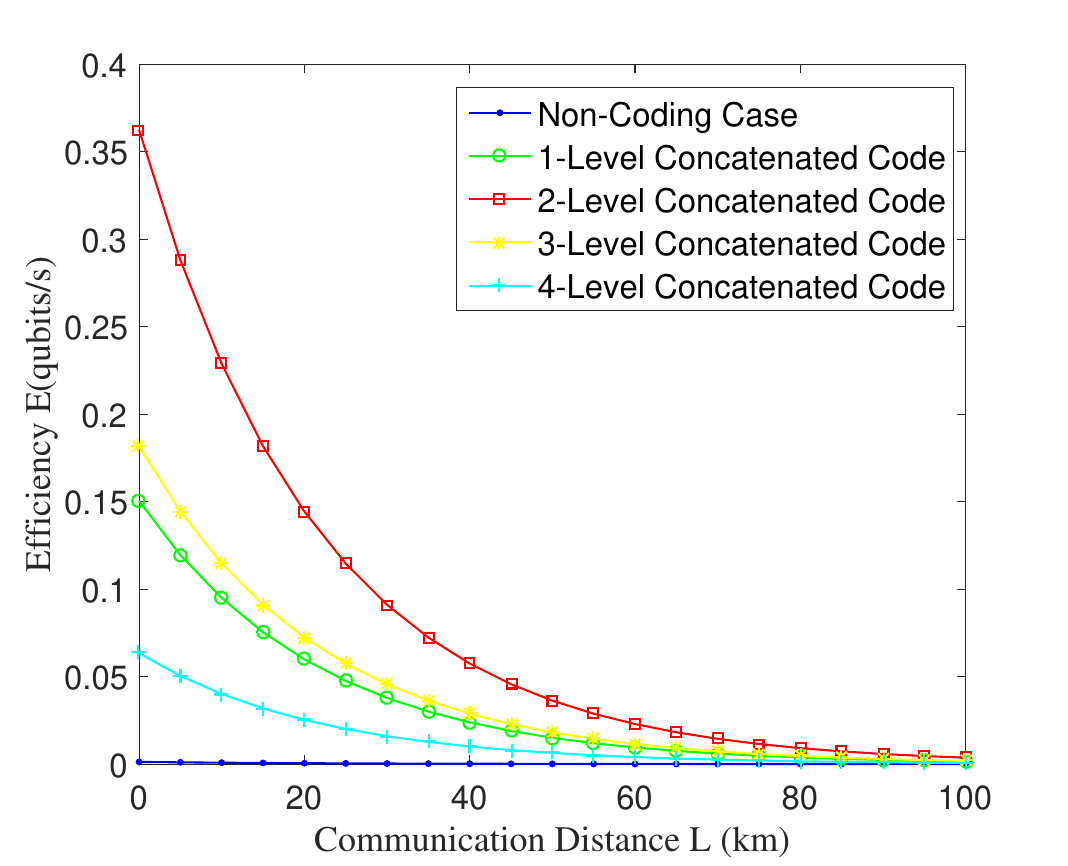}
	\caption{ The preparation efficiency $E$ with the same probability of successful qubit preparation vs. the communication distance $L$. It is the partial enlargement of Fig.\ref{fig-Concatenation-LvsEfficiency}.
 % The figure is a detailed view to show simulation results with 1,2,3,4-level concatenation code and non-coding case in our Protocol.\ref{protocol3-concatenation}. 
 }
	\label{fig-Amplify-Concatenation-LvsEfficiency}
\end{figure}

% \noindent {\bf (d) The efficiency comparison for preparing concatenated [[7,1,3]] codes} 
Fig.\ref{fig-Concatenation-LvsEfficiency} and \ref{fig-Amplify-Concatenation-LvsEfficiency} depict the relationship between the preparation efficiency $E$ and the communication distance $L$, while maintaining the same probability of successful qubit preparation. The green, red, yellow, cyan, and blue lines represent the simulation results of the 1,2,3,4-level concatenation codes and the non-coding case, respectively. The black line corresponds to the simulation results in the asymptotic case which is the infinite data size and with near-perfect qubits preparation. 

In Fig.\ref{fig-Concatenation-LvsEfficiency}, the preparation efficiency $E$ in the asymptotic case is much higher than that of the actual case. Since the fluctuation of finite data and imperfect preparation of required qubits are inevitable in real UBQC, a large number of pulses are used as ancilla qubits to deal with the impacts of fluctuation and qubit errors, which leads to a reduction in efficiency. In order to better show our simulation results, we give a partial enlargement of Fig.\ref{fig-Concatenation-LvsEfficiency}, as shown in Fig.\ref{fig-Amplify-Concatenation-LvsEfficiency}. Not that the preparation efficiency $E$ has a sharp decline trend as the communication distance increases. The error probability of quantum error-correcting code prepared is $e_0^{16}$, and the probability of successful qubit preparation with size $S$ is $(1-e_0^{16})^S$. under the condition of having the same probability of successful qubit preparation, we demonstrate the preparation efficiency of the non-coding case and the concatenation codes with different levels. Although the high-level concatenation code can obtain a lower error rate, it requires massive quantum resource consumption, which results in lower efficiency. Hence, it is not applicable in the reality. In UBQC with limited quantum resources, the 2-level concatenation code is capable of delivering a better preparation efficiency in Fig.\ref{fig-Amplify-Concatenation-LvsEfficiency}.                           

\noindent
{\bf Summary of simulation 3:}
Our Protocol.\ref{protocol3-concatenation} demonstrates that the 2-level concatenation code can be utilized to improve both the preparation efficiency and quantum resource utilization. In addition, Protocol.\ref{protocol3-concatenation} yields significant improvement in the error-correcting performance for fault-tolerant blind quantum computation. These findings highlight the potential of the 2-level concatenation code as an effective method for enhancing the performance and efficiency of UBQC.

% It is also very significant to improve the error-correcting performance in fault-tolerant blind quantum computation.

\section{Related works}
In the original fault-tolerant UBQC~\cite{broadbent2009universal}, fault-tolerant quantum computing is performed on the top of the brickwork state. Thus, the use of CNOT gates acting on non-adjacent qubits in quantum computing requires many SWAP gates for proper implementation, which results in a linear increase in the size of the brickwork state with the scaling of the computation~\cite{broadbent2009universal}. In our Protocol.\ref{protocol2-ftbqc}, the number of ancilla qubits remains relatively constant as the computational size. In addition, when compared to Chien's two fault-tolerant protocols~\cite{chien2015fault}, our protocol only requires Alice to send weak coherent pulses to prepare quantum error-correcting codes, which can free Alice's dependence on quantum computing and quantum memory. In addition, since different quantum gates have different sizes, the used quantum gates can be estimated by Bob when eliminating redundant qubits. However, our Protocol.\ref{protocol2-ftbqc} offers a highly efficient approach for preparing quantum error-correcting codes on cluster state, which not only can reduce quantum resource consumption, but also can ensure that Bob cannot gain any information on the prepared state. Furthermore, our approach supports fault-tolerant quantum computing on encoded brickwork state, which can enhance $\epsilon$ blindness and $\epsilon$ reliability for practical quantum computing applications.

Note that the current quantum technology still struggles with noises and imperfect measurement, which results in a high error rate in the fault-tolerant UBQC. To address this problem, we propose to use the concatenated stabilizer codes to reduce qubit requirements by tailoring circuits to suppress the dominant effect of qubit errors~\cite{guillaud2021error,chamberland2022fault-tolerant}. Chamberland, Jochym, and Laflamme et al. concatenated the 7-qubit Steane code with the 15-qubit Reed-Muller code to implement universal fault-tolerant quantum computation without state distillation~\cite{chamberland2017overhead}. Paul Webster et al. presented a general framework for universal fault-tolerant logic with no-go theorem and stabilizer codes~\cite{webster2022universal}, which can be applied to a wide range of stabilizer code families, including concatenated codes and conventional topological stabilizer codes. Fault-tolerant quantum computation with concatenated quantum codes was proposed by Chamberland, Noh, and Preskill et al.~\cite{chamberland2022fault-tolerant}, which can reduce the consumption of qubits by tailoring the quantum error-correcting codes to suppress the dominant phase-flip errors. To avoid coherent errors, Yingkai Ouyang proposed rotated concatenated stabilizer codes~\cite{ouyang2021concatenated}, namely, concatenating an [[n,k,d]] stabilizer outer code with constant-excitation inner codes. It was shown that when the stabilizer outer code is fault-tolerant, the concatenated codes are immune from coherent phase errors. The concatenation codes above are useful for improving the fault-tolerant threshold and reducing quantum resource consumption in a realistic UBQC system. 

In recent years, many encoding methods~\cite{knill1996concatenated,morimae2012blindtopological,krinner2022realizing,zhang2022concatenation} were proposed to deal with this problem, such as concatenation code~\cite{zhang2022concatenation}, RHG Lattice code~\cite{morimae2012blindtopological} and surface code~\cite{krinner2022realizing}. However, it remains daunting due to the complex encoding structures and large resource consumption. In this paper, we propose a general fault-tolerant blind quantum computation with concatenation codes to reduce the error rate of logical qubits, thereby improving the fault-tolerant threshold of UBQC. In addition, we optimize the number of required pulses in the concatenated code to reduce quantum resource consumption. In fault-tolerant UBQC with limited quantum resources, we showed that the 2-level concatenated code can obtain the optimal performance.  

\section{Conclusions}
In the paper, we first propose a protocol to realize quantum error-correcting codes on cluster state. A fault-tolerant blind quantum computation with quantum error-correcting codes is proposed to address errors in quantum computation. To improve the error-correcting performance, the stabilizer codes are used for multi-level concatenating to improve the fault-tolerant threshold of qubits. To reduce quantum resource consumption, we also analyze the number of required pulses in each level of concatenation code. Since a large number of qubits are required for repetitive encoding in the high-level concatenation codes, the low-level concatenated codes ($0\le n\le4$) are only considered in this paper. We show that the optimal level in the concatenation codes is about $2$, and the resource consumption ratio reaches the minimum value, which means that the number of required pulses in the 2-level concatenation codes is less than other levels under the same probability of successful qubit preparation. From the perspective of reducing quantum resources, our simulation results show that the 2-level concatenated code can achieve a better performance than other levels with the increase of communication distance. To further improve the practical performance of UBQC, our future works include designing other concatenated codes, stabilizer codes, and surface codes. At the same time, we also plan to use topological codes with high fault tolerance features to perform distributed blind quantum computation with limited quantum resources. In summary, our encoded scheme has the potential to enhance the reliability and blindness of blind quantum computation protocols with different graphs, making it a versatile and practical solution for a wide range of quantum computing applications.

\bibliographystyle{IEEEtran}
\begin{small}
	\bibliography{UBQC_reference}
\end{small}

% \appendices

% \section{}

% % use section* for acknowledgment
% \ifCLASSOPTIONcompsoc
%   % The Computer Society usually uses the plural form
%   \section*{Acknowledgments}
% \else
%   % regular IEEE prefers the singular form
% \fi


\ifCLASSOPTIONcaptionsoff
  \newpage
\fi

\end{document}